\documentclass[a4paper,11pt]{article}
\pdfoutput=1

\usepackage[T1]{fontenc}
\usepackage{aas_macros}
\usepackage{jcappub}

\usepackage{amsmath,amssymb,mathtools}
\usepackage{physics}
\usepackage{braket}
\usepackage{graphicx}
\usepackage{xcolor}
\usepackage{comment}

\usepackage{subfigure}

\usepackage{tikz}
\usetikzlibrary{matrix,fit,positioning,arrows.meta,decorations.pathreplacing,shapes.geometric,backgrounds}

\newcommand{\bk}{\mathbf{k}}
\newcommand{\Hess}{\mathcal{H}}
\newcommand{\Oop}{\mathcal{O}}

\newcommand{\avg}[1]{\left\langle #1\right\rangle}

\title{\boldmath Spectral Hierarchy of the Cosmic Web}

\author[1,2]{Francisco-Shu Kitaura}
\author[3,4,5,6,1,2]{and Francesco Sinigaglia}
\affiliation[1]{Instituto de Astrof\'{\i}sica de Canarias, s/n, E-38205, La Laguna, Tenerife, Spain}  
\affiliation[2]{Departamento de Astrof\'{\i}sica, Universidad de La Laguna, E-38206, La Laguna, Tenerife, Spain}
\affiliation[3]{Institute for Fundamental Physics of the Universe, Via Beirut 2, I-34151 Trieste, Italy}
\affiliation[4]{SISSA - International School for Advanced Studies, Via Bonomea 265, 34136 Trieste, Italy}
\affiliation[5]{INAF - Osservatorio Astronomico di Trieste, Via G. B. Tiepolo 11, I-34131 Trieste, Italy}
\affiliation[6]{INFN – National Institute for Nuclear Physics, Via Valerio 2, I-34127 Trieste, Italy}

\emailAdd{fkitaura@iac.es}

\abstract{
We introduce a {spectral hierarchy} of cosmic-web classifications obtained by applying simple scale-weighting kernels to the density field before performing a standard eigenvalue-based web classification.
This unifies and extends several widely used web definitions within a single framework: the familiar potential/tidal web (large-scale, nonlocal), a curvature-based web (more local, peak- and ridge-sensitive), and additional higher-derivative levels that progressively emphasize smaller-scale structure.
Because the classification is built from second derivatives of the filtered field, successive hierarchy levels align naturally with operator families that appear in renormalised bias and effective descriptions of large-scale structure, providing an explicit bridge between cosmic-web environments and long- and short-range nonlocal bias ingredients.

We quantify the information content of the hierarchy with a compact statistic: we map each cell to one of four ordered web types (void, sheet, filament, knot), construct a corresponding ``web contrast'' field, and measure its cross-correlation with halos from the \textsc{AbacusSummit} simulation suite on a coarse mesh with $\Delta L\simeq 5.5\,h^{-1}\mathrm{Mpc}$.
We find that the hierarchy retains significant tracer-relevant information from very large scales down to the mesh Nyquist limit, with the more local (curvature/higher-derivative) levels dominating toward nonlinear scales.
This makes the spectral hierarchy a practical, interpretable conditioning basis for fast mock-galaxy production and field-level modelling, and a flexible tool for studying environment-dependent clustering and assembly bias.
}

\begin{document}
\maketitle
\flushbottom

%=========================================================
\section{Introduction}
\label{sec:intro}

The cosmic web --- the network of voids, sheets, filaments and knots --- is a defining feature of the late-time Universe
\citep[e.g.,][]{2008LNP...740..335V,2009LNP...665..291V,2018MNRAS.473.1195L}.
It encodes anisotropic gravitational collapse and provides a natural language for describing environment-dependent phenomena,
including galaxy formation pathways and assembly bias.
Cosmic web information is exploited across a broad range of applications:
missing baryons in filaments, void cosmology, redshift-space distortions, dark energy constraints, and tests of gravity
\citep[e.g.,][]{2015Natur.528..105E,2016PhRvL.116q1301K,2017PhRvD..95f3528C,2019MNRAS.484.5267K}.
A long-standing challenge is that ``cosmic web'' remains method-dependent; different classifiers emphasize different aspects of the field
\citep{2018MNRAS.473.1195L}.

In parallel, modern large-scale structure analysis increasingly relies on forward models and fast mock catalogues.
Accurate mock generation for current and upcoming surveys requires an efficient yet physically grounded representation of
tracer bias, including nonlocality and stochasticity.
Effective field-level bias models have matured significantly \citep{McDonald09,Desjacques_2018},
and a key insight is that nonlocal bias operators are closely related to tidal and curvature information in the underlying density field.
A particularly powerful approach is the hierarchical cosmic web assembly-bias framework,
which treats different cosmic web regions as distinct tracer subsets and thereby captures long- and short-range nonlocality indirectly
\citep{Kitaura_2022,Coloma_2024}.

In this work we extend the concept of cosmic web classification by introducing a {spectral hierarchy}
defined by kernel operators acting on the density contrast in Fourier space.
This hierarchy:
(i) unifies standard potential-based and curvature-based web definitions as particular levels,
(ii) naturally connects to long- and short-range nonlocal bias operators,
(iii) offers additional levels motivated by higher-derivative bias and effective field theory,
and (iv) yields a compact information-theoretic compression of environment that demonstrably correlates with halo catalogues down to mesh Nyquist scales relevant for fast mock generation.

{The present work is also a companion paper to a series of studies that use the \texttt{WebOn} framework—most notably its spectral hierarchical cosmic-web classification—as a core ingredient for field-level bias modelling and subgrid tracer placement \citep[e.g.][]{Coloma_2024,Forero_2024}.}

%=========================================================
\section{The cosmic web framework and generalized kernel hierarchy}
\label{sec:cw}

In this section we recap the foundations of cosmic web classification and explain why it provides a natural language to design
{scale-separated} and {hierarchical} operators for large-scale structure analyses and fast forward modelling.
The cosmic web is the emergent anisotropic pattern of knots, filaments, sheets and voids produced by gravitational instability.
A broad class of cosmic web classifiers can be understood as diagnostics of {second derivatives} of either the gravitational
potential or the density field, i.e.\ of the local curvature of the evolved large-scale structure.

Standard (tidal) cosmic web approaches are based on the tidal tensor,
\begin{equation}
T_{ij}(\mathbf{x}) \equiv \partial_i\partial_j \phi(\mathbf{x}),
\label{eq:tidal_tensor}
\end{equation}
where $\phi$ is the { rescaled peculiar gravitational potential} obeying $\nabla^2\phi(\mathbf{x})=\delta(\mathbf{x})$.
The eigenvalues of $T_{ij}$ quantify collapse/expansion along principal directions, motivating the classification of the environment
into four morphological components (knots, filaments, sheets and voids) according to the number of eigenvalues above a threshold.

{This construction is rooted in the local equations of motion. 
For a test particle in comoving coordinates, the peculiar gravitational acceleration is
\begin{equation}
{ \ddot{x}_i+2H\dot{x}_i = -4\pi G\bar\rho_m\,\partial_i \phi(\mathbf{x}),}
\end{equation}
where dots denote derivatives with respect to { cosmic time, $H$ is the Hubble rate, and $\phi$ is normalised by $\nabla^2\phi=\delta$}.
Expanding the force to first order around a reference position $\bar{\mathbf{x}}$ gives the linearized equation for the separation
$\Delta x_i \equiv x_i-\bar{x}_i$ of neighbouring fluid elements,
\begin{equation}
{ \Delta\ddot{x}_i+2H\Delta\dot{x}_i
= -4\pi G\bar\rho_m \sum_j T_{ij}(\bar{\mathbf{x}})\,\Delta x_j .}
\label{eq:linear_eom_tidal}
\end{equation}
Thus the tidal tensor governs the relative acceleration along the principal axes of the flow.
After diagonalization, Eq.~\eqref{eq:linear_eom_tidal} reduces locally to three independent equations,
\begin{equation}
{ \Delta\ddot{x}_i+2H\Delta\dot{x}_i = -4\pi G\bar\rho_m\lambda_i\Delta x_i ,}
\end{equation}
where $\lambda_i$ are the eigenvalues of $T_{ij}$.
{ Their signs determine whether the relative gravitational acceleration is compressive or extensive along each principal direction}, providing the dynamical basis for classifying regions into knots, filaments, sheets, and voids, as emphasized in Hahn et al. 2007 \citep{2007MNRAS.375..489H}.

While the tidal tensor provides the most direct dynamical description of anisotropic collapse, it is only one particular realization of a broader second-derivative perspective on the cosmic web.
A closely related and more local construction is obtained by acting directly on the evolved density field rather than on the gravitational potential.
This leads to classifiers built from the Hessian of the density,
\begin{equation}
H_{ij}(\mathbf{x}) \equiv \partial_i\partial_j \delta(\mathbf{x}),
\label{eq:density_hessian}
\end{equation}
which are more sensitive to short-range curvature and peak-like structure, and therefore provide a complementary handle on the small-scale geometry of the web.
The tidal tensor and the density Hessian can thus be regarded as adjacent members of a wider family of anisotropy diagnostics: the former probes non-local, potential-driven structure formation, whereas the latter probes the local curvature of the matter field itself.
In both cases, the eigenvalues and rotational invariants capture anisotropic information beyond the local density and provide a natural bridge between cosmic-web morphology and operator-based descriptions of large-scale structure.

}

Motivated by this, we introduce the notion of a {generalized kernel hierarchy}:
a family of kernels defining progressively different degrees of nonlocality and derivative order that can be applied either
(i) as classifiers (discrete web labels) or
(ii) as continuous fields (filtered curvature and ridge measures),
with the goal of organizing information content across scales in a controlled and computationally efficient way.
In particular, by arranging operators in a spectral ladder, we can interpret web information as a compact summary of
the field content that enters bias expansions and effective stress-tensor descriptions.

% ---------------------------------------------------------
\subsection{Lagrangian--Eulerian mapping and the role of second derivatives}
\label{sec:cw_mapping}

The Lagrangian--Eulerian mapping,
\begin{equation}
\mathbf{x}(\mathbf{q},z)=\mathbf{q}+\mathbf{\Psi}(\mathbf{q},z),
\label{eq:xe_map_cw}
\end{equation}
encodes structure formation through the displacement field $\mathbf{\Psi}$.
The Eulerian density contrast can be written in terms of the Jacobian of this mapping,
\begin{equation}
1+\delta(\mathbf{x})
=
\sum_{\mathbf{q}_*}
\left|
\det\left(
\delta_{ij}^{\rm K}
+
\frac{\partial \Psi_i}{\partial q_j}
\right)
\right|^{-1}_{\mathbf{q}_*}\,,
\label{eq:jacobian_density}
\end{equation}
where the sum runs over all streams $\mathbf{q}_*$ mapping to the same Eulerian position.
Before shell crossing, the mapping is single-valued and local collapse is controlled by the eigenvalues of the deformation tensor
$\partial\Psi_i/\partial q_j$.
After shell crossing, multi-streaming appears and the density field develops sharp anisotropic features (ridges).

Second derivatives enter naturally at two connected levels:

\paragraph{(i) Large-scale anisotropic transport.}
In perturbation theory, the displacement is potential, $\mathbf{\Psi}=-\nabla\Phi$, so the deformation tensor is a Hessian.
In the Zel'dovich approximation and its higher-order extensions, tidal effects are therefore captured by second derivatives
of a potential, and the eigen-structure of such Hessians determines preferential collapse directions.

\paragraph{(ii) Eulerian curvature and ridge sharpening.}
Once particles have been transported to Eulerian space, the local morphology of the density field---whether matter lies in a knot,
filament, sheet or void---is governed by the eigenvalues of curvature tensors such as $T_{ij}$ and $H_{ij}$.
Ridges correspond to directions of maximal compression and large curvature transverse to the ridge.
This is the regime where higher-derivative operator content becomes increasingly relevant: it characterizes how short-range morphology
reacts to coarse-graining and how additional relaxation can be parameterized in effective descriptions.

\paragraph{Generalized kernel hierarchy.}
To make this connection explicit, consider a family of kernels defining { a family of transformed fields},
\begin{equation}
\delta^{(n)}(\mathbf{k}) \equiv K_n(k)\,\hat{\delta}(\mathbf{k}).
\label{eq:kernel_hierarchy}
\end{equation}
{ Here $n$ labels the operator level; no pointwise ordering of the kernels is assumed. The analysis below uses even powers of $k$, not Gaussian filters, with increasing order enhancing relative sensitivity to short-scale variations.}
One can then define hierarchical curvature tensors,
\begin{equation}
T^{(n)}_{ij}(\mathbf{x}) \equiv \partial_i\partial_j \phi^{(n)}(\mathbf{x}),
\qquad
H^{(n)}_{ij}(\mathbf{x}) \equiv \partial_i\partial_j \delta^{(n)}(\mathbf{x}),
\label{eq:hier_curvature}
\end{equation}
and their eigenvalues/invariants to obtain either discrete web labels at each level $n$ or continuous ridge measures.
The spectral hierarchy introduced below provides an analytically simple and operator-aligned version of this idea, in which
the filtration is organized by integer even powers of $k$.

%=========================================================
\subsection{Spectral kernel hierarchy}
\label{sec:spectral_hierarchy}

We define a family of filtered fields via Fourier-space kernel operators
\begin{equation}
\widehat{\delta}^{(i)}(\mathbf{k}) \equiv -k^i\,\widehat{\delta}(\mathbf{k}),\qquad i\in 2\mathbb{Z}.
\label{eq:kernel_def}
\end{equation}
{ The overall minus sign is a convention consistent with potential-based operators (e.g., { $\widehat\phi=-k^{-2}\widehat\delta$}), but it also ensures a common eigenvalue interpretation across the hierarchy. For example, at $i=0$ one has $\delta^{(0)}=-\delta$ and $\Hess^{(0)}_{ij}=-\partial_i\partial_j\delta$. At an ideal density maximum the Hessian of $\delta$ is negative definite, so the Hessian of $-\delta$ has positive eigenvalues. Consequently, the usual convention in which a knot corresponds to three eigenvalues above the threshold can be applied consistently to both potential-based and density-curvature levels.}
In real space, the index $i$ controls the degree of derivative/nonlocality:
\begin{equation}
{ \delta^{(i)}(\mathbf{x})=(-1)^{i/2+1}(\nabla^2)^{i/2}\delta(\mathbf{x}),\qquad i\in2\mathbb Z.}
\end{equation}

{ Negative powers denote inverse Laplacians on nonzero Fourier modes.}

For each level $i$, we define the generalized Hessian tensor
\begin{equation}
\Hess^{(i)}_{ab}(\mathbf{x}) \equiv \partial_a\partial_b \delta^{(i)}(\mathbf{x})\,.
\label{eq:general_hessian}
\end{equation}

Interestingly, the spectral hierarchy reproduces the same derivative ladder
that appears in renormalized bias and EFT descriptions of large-scale
structure. 
Because the classification tensor involves two additional
derivatives of the filtered field, its scalar invariants involve
operator structures scaling as $\nabla^2\delta$, $\nabla^4\delta$,
$\nabla^6\delta$, and higher orders, corresponding to the gradient expansion discussed
in bias theory and effective field theory
\citep{McDonald09,Carrasco_2012,Desjacques_2018}.

%=========================================================
\section{Physical interpretation of hierarchy levels}
\label{sec:phys}

The generalized kernel hierarchy introduced in Sec.~\ref{sec:cw} provides a unified way of organizing gravitational evolution and effective corrections according to their derivative order.
Each hierarchy level can be associated with a characteristic physical regime, ranging from relativistic large-scale effects to highly nonlinear short-range relaxation.
In this section we provide a physical interpretation of the different levels in terms of scale, dynamics, and their relevance for fast forward models.

We will use a compact operator notation throughout. For even $i$ we define
\begin{equation}
\Oop_i[{\delta}](\bk)\equiv -k^i\,\hat{\delta}(\bk),
\qquad i\in 2\mathbb{Z},
\label{eq:Oi_def}
\end{equation}
so that $\delta^{(i)}=\Oop_i[\delta]$ and $\Hess^{(i)}_{ab}=\partial_a\partial_b\,\Oop_i[\delta]$.
Negative indices correspond to inverse-Laplacian (long-range) operators and positive indices to increasingly local, short-range operators.
Long-range levels encode coherent gravitational transport and tidal evolution,
whereas higher positive levels encode local relaxation, curvature sensitivity, and small-scale stabilization.
The intermediate levels correspond to the standard Newtonian regime captured by LPT and its augmented variants.
A compact summary of the levels is provided in Tab.~\ref{tab:kernels}.

\subsection{$i=-4$: relativistic/gauge-screening sensitivity}

The level $i=-4$ corresponds to operators involving higher inverse powers of the Laplacian,
schematically of the form
\begin{equation}
\Oop_{-4}[\delta]\sim \nabla^{-4}\delta.
\end{equation}
Such terms are extremely long-range and sensitive to large-scale boundary conditions and gauge choices \citep[see][]{2016PhRvD..94h3511H}.
Physically, they can be interpreted as encoding relativistic and screening-like effects that become relevant on horizon scales.

In Newtonian gravity, the dominant long-range operator is $\nabla^{-2}\delta$, which gives rise to the Poisson equation.
However, when considering relativistic corrections or effective large-scale modifications (e.g. gauge effects in cosmological perturbation theory, infrared resummations, or horizon-scale screening),
higher inverse-Laplacian structures naturally appear.
These operators amplify the contribution of ultra-large-scale modes and therefore control the coupling between super-horizon perturbations and sub-horizon evolution.

{Relativistic corrections introduce an additional infrared contribution
to the displacement field that depends on a time-dependent screening
length $\ell(a)$ \citep{2016PhRvD..94h3511H}.
Using the approximation derived in that work, the shift between the conformal 
Newtonian (CN) gauge particle positions and the simulation comoving synchronous (CS) coordinates can
be written in Fourier space as
\begin{equation}
\mathbf{\Psi}_{\rm CS\rightarrow CN}(\mathbf{k},a)
\approx
\frac{i\mathbf{k}}{\ell^2(a)\,k^4}\,\delta(\mathbf{k},a),
\end{equation}
so that the divergence of the displacement correction becomes
\begin{equation}
\nabla\cdot\mathbf{\Psi}_{\rm CS\rightarrow CN}
=
{ -\frac{1}{\ell^2(a)\,k^2}\,\delta(\mathbf{k},a).}
\end{equation}
The time-dependent screening length is approximately
\begin{equation}
\ell(a) =
\ell_0\, a^{2/7}\left(\frac{H}{H_0}\right)^{-3/7},
\qquad
\ell_0=\frac{c}{\sqrt{3}H_0}\Omega_m^{-2/7}.
\end{equation}
This expression shows that relativistic corrections introduce an
{ $k^{-2}$ contribution to the displacement divergence, generated by a scalar potential proportional to $k^{-4}\delta$ and therefore associated with the $i=-4$ scalar level}, while the factor $\ell(a)$ provides a horizon-scale
screening that suppresses the effect on sub-horizon modes \citep[see also][]{Bernardeau_2002,2011PhRvD..83l3505C,2011JCAP...10..031B,2013PhRvD..88j3527A,2016NatPh..12..346A}.
}

In the context of fast gravity solvers, $i=-4$ levels are typically subdominant for mock catalogue production at sub-horizon scales.
Nevertheless, they provide a formal link between Newtonian operator hierarchies and relativistic perturbation theory.
In particular, they highlight that the lowest hierarchy levels are sensitive to the global cosmological frame and to long-wavelength modes that modulate local structure formation (the separate-universe picture).
From a practical perspective, $i=-4$ contributions may become relevant when generating very large-volume lightcones approaching horizon scales or when incorporating relativistic light-cone effects and gauge corrections.

Primordial non-Gaussianity (PNG) provides another example of physically
relevant scale-dependent structure on the infrared side of the hierarchy.
In particular, local-type PNG induces the well-known large-scale
scale-dependent halo-bias correction
\begin{equation}
\Delta b(k)\propto f_{\rm NL}\,k^{-2}\,,
\end{equation}
{ which corresponds to the trace of the $i=-4$ classification tensor: $\widehat{\mathrm{Tr}\,\Hess^{(-4)}}=k^{-2}\widehat\delta$. The index labels the scalar field before taking its Hessian, which adds two derivatives \citep{Dalal_2008,Desjacques_2018}.}
PNG is commonly parameterized through the dimensionless coefficient
$f_{\rm NL}$ by writing the primordial Bardeen potential as
\begin{equation}
\Phi(\mathbf{q})=
\phi(\mathbf{q})
+
f_{\rm NL}\left[\phi^2(\mathbf{q})-\langle\phi^2\rangle\right],
\end{equation}
where $\phi$ is a Gaussian random field in Lagrangian coordinates
$\mathbf{q}$  \citep[][for relevant studies]{Dalal_2008, GRPNG1,BIASPNG1,2012PhRvD..85d1301B,2011JCAP...10..031B}.

{ The quadratic primordial coupling produces a leading large-scale tracer response proportional to $f_{\rm NL}\Phi$, giving $\Delta b\propto f_{\rm NL}/[k^2T(k)D(z)]$. Thus the $k^{-2}$ correspondence applies as $T(k)\to1$. It identifies the infrared operator scaling, not a complete PNG model obtained by applying the classifier to a nonlinear density field.}

\subsection{$i=-2$: Newtonian gravitational potential}

The level $i=-2$ corresponds to the standard Newtonian Poisson operator,
\begin{equation}
\Oop_{-2}[\delta]\sim \nabla^{-2}\delta,
\end{equation}
which determines the gravitational potential through $\nabla^2\phi \propto \delta$.
This level defines the backbone of structure formation in the Newtonian regime.
It encodes the long-range, non-local nature of gravity: density perturbations source a potential that extends over arbitrarily large distances.
In Fourier space, the operator scales as $k^{-2}$, implying that large-scale modes dominate the gravitational response.

All Lagrangian Perturbation Theory (LPT) approaches---including 2LPT and ALPT---are fundamentally built upon this level.
The displacement field,
\begin{equation}
\mathbf{\Psi} \sim -\nabla \nabla^{-2}\delta,
\end{equation}
is directly constructed from $\Oop_{-2}$.
Thus, $i=-2$ governs coherent bulk flows, tidal transport, and large-scale anisotropic collapse, providing the correct infrared behaviour required for large-volume mock production.

{The same tidal tensor that governs large-scale anisotropic collapse also plays a central role in tidal torque theory (TTT), which explains the origin of angular momentum in proto-haloes \citep{Peebles_1969,1970Afz.....6..581D,Catelan_Theuns_1996,Heavens_peacock_1988,Porciani_2002a,Porciani_2002b}.
In this framework the angular momentum of a collapsing proto-halo is generated by the misalignment between its inertia tensor and the surrounding tidal field $T_{ij}=\partial_i\partial_j\phi$.
At leading order the torque scales schematically as
\begin{equation}
L_i \propto \epsilon_{ijk}\,I_{jl}\,T_{lk},
\end{equation}
where $I_{ij}$ is the inertia tensor of the proto-halo region.
Since the tidal tensor corresponds precisely to the $i=-2$ level of the spectral hierarchy, this level also encodes the large-scale tidal environment responsible for generating halo angular momentum.
In this sense, the potential-based cosmic web classification naturally captures the large-scale dynamical structures that influence both anisotropic collapse and the spin acquisition of dark matter haloes.
}

\subsection{$i=0$: curvature web and short-range nonlocal bias}

The level $i=0$ corresponds to the operator: 
\begin{equation}
\Oop_{0}[\delta]\sim -\delta,
\end{equation}
and
\begin{equation}
\Hess^{(0)}_{ij}=-\partial_i\partial_j\delta.
\end{equation}

In Fourier space this tensor scales as $k_i k_j\,\delta(\bk)$,
so its trace corresponds to

\begin{equation}
\mathrm{Tr}\,\Hess^{(0)} \propto \nabla^2\delta.
\end{equation}

Thus the $i=0$ level is directly connected to the leading higher-derivative bias operator $\nabla^2\delta$ discussed in {peak theory} \citep[][]{Bardeen_1986,Desjacques_2010} and renormalized bias theory \citep[e.g.,][]{McDonald09,Desjacques_2018}.
Physically, this operator encodes finite Lagrangian size and formation-scale sensitivity of tracers.
It represents the first short-range nonlocal correction beyond density and tidal invariants.

In geometric terms, $i=0$ defines a curvature web:
regions are classified according to the Hessian of the density field,
emphasizing peaks, ridges and saddle morphology \citep[][]{Heavens_peacock_1988,Sinigaglia_2021}.

\subsection{$i=+2$: second higher-derivative level}

For $i=2$ one obtains
\begin{equation}
\delta^{(2)}(\bk)=-k^2\delta(\bk),
\end{equation}
so that in real space (up to the overall sign convention)
\begin{equation}
\Oop_{2}[\delta]\sim\delta^{(2)}(\mathbf{x}) \propto \nabla^2\delta(\mathbf{x}).
\end{equation}
The generalized classification tensor is
\begin{equation}
\Hess^{(2)}_{ij} \equiv \partial_i\partial_j \delta^{(2)}(\mathbf{x})
\quad \Longleftrightarrow \quad
\Hess^{(2)}_{ij}(\bk)\propto k_i k_j k^2 \delta(\bk).
\end{equation}
Accordingly, the leading rotationally invariant scalar built from the classification tensor (its trace) scales as
\begin{equation}
\mathrm{Tr}\,\Hess^{(2)}=\nabla^2\delta^{(2)} \propto \nabla^4\delta.
\end{equation}
Such operators appear naturally as subleading higher-derivative terms in bias expansions
and as next-order gradient corrections in effective field theory descriptions of large-scale structure.
Physically, they encode sensitivity to curvature gradients
and to increasingly localized morphological structure.
Their contribution becomes relevant toward smaller scales where
finite-size and coarse-graining effects are more pronounced.

\subsection{$i=+4$: third higher-derivative level}

For $i=4$ one has
\begin{equation}
\delta^{(4)}(\bk)=-k^4\delta(\bk),
\end{equation}
so that in real space (again up to the overall sign convention)
\begin{equation}
\Oop_{4}[\delta]\sim\delta^{(4)}(\mathbf{x}) \propto \nabla^4\delta(\mathbf{x}).
\end{equation}
The corresponding classification tensor satisfies
\begin{equation}
\Hess^{(4)}_{ij}(\bk)\propto k_i k_j k^4 \delta(\bk),
\end{equation}
and its trace scales as
\begin{equation}
\mathrm{Tr}\,\Hess^{(4)}=\nabla^2\delta^{(4)} \propto \nabla^6\delta.
\end{equation}
Operators of this type arise as higher-order UV-sensitive terms
in renormalized bias theory and in effective field theory treatments,
where successive gradient corrections encode increasingly short-range response.
While typically suppressed at large scales,
they become relevant near the ultraviolet cutoff of a numerical mesh
or in highly nonlinear environments.

%=========================================================
\begin{table}[t]
\centering
\small
\begin{tabular}{|c|c|p{9.6cm}|}
\hline
\textbf{Level $i$} & \textbf{Field $\delta^{(i)}(\bk)$} & \textbf{Physical interpretation / operator content} \\
\hline
$-4$ & $-k^{-4}\,\hat{\delta}(\bk)$ &
\textbf{IR-enhanced relativistic/gauge-screening level.}
Real-space $\delta^{(-4)}\propto -\nabla^{-4}\delta$.
Connects to conformal Newtonian gauge GR-screening/displacement-like corrections; emphasizes ultra-large-scale morphology; { its Hessian trace also has the large-scale $k^{-2}\delta$ scaling of local-PNG bias \citep{Dalal_2008}} \citep{Bernardeau_2002,2011PhRvD..83l3505C,2011JCAP...10..031B,2013PhRvD..88j3527A,2016NatPh..12..346A,2016PhRvD..94h3511H}. \\
\hline
$-2$ & $-k^{-2}\,\hat{\delta}(\bk)$ &
\textbf{Potential / tidal-tensor level (classical T-web).}
$\delta^{(-2)}\propto \phi$ with $\nabla^2\phi=\delta$ and
$\Hess^{(-2)}_{ij}=\partial_i\partial_j\phi$.
Encodes anisotropic collapse and long-range tidal environments; relates to long-range nonlocal bias through tidal invariants (e.g., $s^2,s^3$) \citep[][]{Zeldovich_1970,1970Afz.....6..581D,White_1984,Porciani_2002a,Porciani_2002b,Hahn_2007,Heavens_peacock_1988}. \\
\hline
$0$ & $-k^{0}\,\hat{\delta}(\bk)$ &
\textbf{Curvature level (density Hessian).}
$\delta^{(0)}=-\delta$ and $\Hess^{(0)}_{ij}=-\partial_i\partial_j\delta$.
The trace gives $\mathrm{Tr}\,\Hess^{(0)}\propto \nabla^2\delta$.
Highlights curvature/peak morphology; connects to tidal-torque ingredients and short-range conditioning used in bias models \citep{Heavens_peacock_1988,McDonald09,Desjacques_2018,Coloma_2024}. \\
\hline
$2$ & $-k^{2}\,\hat{\delta}(\bk)$ &
\textbf{Next higher-derivative level.}
$\delta^{(2)}\propto \nabla^2\delta$, but the web classifier uses a Hessian, so its scalar invariants carry two additional derivatives; in particular
$\mathrm{Tr}\,\Hess^{(2)}\propto \nabla^4\delta$.
This is the next derivative order {in the invariants}, and it becomes relevant toward the UV/mesh cutoff. Related to higher-derivative bias and EFT operator expansions; can also
provide an effective language for scale-dependent responses arising from
massive neutrinos, non-standard dark matter, or modified-gravity models. \citep{2006PhR...429..307L,Carrasco_2012,2013JCAP...08..037P,Senatore_2015,Senatore_2015b,Desjacques_2018,Tsujikawa_2007,DeFelice_Tsujikawa_2010}.  \\
\hline
$4$ & $-k^{4}\,\hat{\delta}(\bk)$ &
{\textbf{Highest UV / curvature-gradient level considered here.}
$\delta^{(4)}\propto\nabla^4\delta$, while the Hessian classifier contains two additional derivatives, so that
$\mathrm{Tr}\,\Hess^{(4)}\propto\nabla^6\delta$.
This level probes extremely localized curvature variations and represents the next term in the higher-derivative expansion beyond the $i=2$ level. It is therefore especially sensitive to structures close to the resolution or coarse-graining scale, where higher-order EFT/bias counterterms, finite-size responses, numerical smoothing, and discretisation effects become increasingly important. Because of this enhanced UV sensitivity, the $i=4$ classification is also more susceptible to mesh-scale noise and may require explicit smoothing or threshold regularisation.} \\
\hline
\end{tabular}
\caption{Spectral hierarchy levels for the kernel ladder $\delta^{(i)}(\bk)\equiv -k^i\,\hat{\delta}(\bk)$ (even $i$). Each level corresponds to a distinct degree of nonlocality/derivative order and aligns with operator sectors used in renormalised bias and EFT descriptions. { Throughout the paper, the hierarchy label $i$ refers to the filtered scalar field $\delta^{(i)}$; the Hessian classifier carries two additional spatial derivatives.}}
\label{tab:kernels}
\end{table}

\subsection{$i=+2$ and $i=+4$:  connection to cosmological physics}

Scale-dependent operators similar to those appearing in the ultraviolet
levels of the spectral hierarchy also arise in cosmological models with
additional particle species or modified dark-matter dynamics.
From the perspective of the spectral hierarchy, such behaviour can be viewed
as motivating effective scale-dependent responses that, in an appropriate
gradient expansion, are analogous to the operator structures associated with
the $i=0$ and $i=2$ levels.
%For example, massive neutrinos introduce a scale-dependent suppression of structure growth due to free-streaming.
In the linear regime this effect can be described through a modified
growth response that depends on $k^2/k_{\rm fs}^2$ relative to the
neutrino free-streaming scale $k_{\rm fs}$ \citep[][]{2006PhR...429..307L}.

From the perspective of the spectral hierarchy, such behaviour
corresponds to a response that depends on higher spatial derivatives
of the density field, analogous to the operators associated with the
$i=0$ and $i=2$ levels.

Similarly, several dark-matter models introduce characteristic
scale-dependent corrections to the effective gravitational response.
Examples include warm dark matter, fuzzy/ ultralight dark matter \citep[][]{2000PhRvL..85.1158H},
and effective fluid descriptions of small-scale dark matter dynamics \citep[][]{Carrasco_2012,2013JCAP...08..037P,Senatore_2015,Senatore_2015b} and bias expansions \citep[][]{Desjacques_2018}.
These models typically produce gradient terms in the effective Euler
equation or modified Poisson relations involving operators such as
$\nabla^2\delta$ or $\nabla^4\delta$.
Such terms correspond naturally to the higher rungs ($i=2,4$) of the
spectral hierarchy and therefore provide a physical interpretation of
these levels in terms of short-range dynamical corrections.
While the present work does not attempt to model these effects
explicitly, the hierarchy provides a convenient language for organizing
and potentially detecting scale-dependent signatures associated with
neutrino free-streaming or non-standard dark-matter physics \citep[see][]{2024A&A...690A..27G}.

{A useful complementary perspective on the ultraviolet rungs of the hierarchy is also provided by modified-gravity. In $f(R)$ gravity the gravitational action is modified by replacing the Ricci scalar $R$ with a function $f(R)$, so that cosmic acceleration can arise from modified spacetime geometry rather than from a cosmological constant.
For example, in the quasistatic limit of $f(R)$ gravity the effective gravitational constant becomes scale dependent and takes the form
\begin{equation}
G_{\rm eff}(k,a)={ \frac{G}{F}}\,
\frac{1+4\,m\,k^2/(a^2R)}{1+3\,m\,k^2/(a^2R)}\,,
\end{equation}
where $F\equiv \partial f/\partial R$ and
\begin{equation}
m \equiv \frac{{\rm d}\ln F}{{\rm d}\ln R}
\end{equation}
controls the scale dependence of the modified response \citep{Tsujikawa_2007,DeFelice_Tsujikawa_2010}.
Expanding in the small parameter
\begin{equation}
x\equiv \frac{m\,k^2}{a^2R}\ll 1
\end{equation}
yields
\begin{equation}
\frac{1+4x}{1+3x}
=
1+x-3x^2+9x^3+\cdots.
\end{equation}
The modified Poisson equation can then be written schematically in configuration space as
\begin{equation}
\nabla^2\Phi
=
4\pi G a^2 \bar\rho_m\,
\frac{1}{F}
\left[
\delta
-\frac{m}{a^2R}\nabla^2\delta
-\frac{3m^2}{(a^2R)^2}\nabla^4\delta
-\frac{9m^3}{(a^2R)^3}\nabla^6\delta
+\cdots
\right].
\end{equation}
This makes explicit that higher-derivative operators such as $\nabla^2\delta$, $\nabla^4\delta$ and $\nabla^6\delta$ arise naturally in a controlled expansion of the scale-dependent gravitational response.
A closely related logic appears in EFT-based descriptions of large-scale structure, where successive higher-derivative terms encode the effect of unresolved short-scale physics and finite-size response.
}

%=========================================================
\section{Effective Euler equation and late-time scale-dependent growth}
\label{sec:effective_growth}

{ This section is a phenomenological illustration; the modified growth equation is not used in the simulations or in the classification and clustering analyses below.}

In two-step or relaxation-augmented gravity solvers, the short-range step can be interpreted
as an additional effective contribution to the matter Euler equation beyond pressureless dust.
At the field level, such corrections arise from coarse-graining unresolved small-scale dynamics and can be organised as a gradient expansion
of an effective stress tensor in powers of spatial derivatives of the density contrast.
To leading order, this produces operators of the form $\nabla^2\delta$ and $\nabla^4\delta$.

These operators do not represent primordial modifications of gravity.
Rather, they encode late-time, small-scale response effects associated with halo relaxation, virialisation, and unresolved nonlinear collapse.
In the context of the spectral hierarchy, they provide a continuous description of the ultraviolet levels ($i=2,4$)
as controlled, scale-dependent modifications of growth.
Higher-order tensors such as
$\partial_i\partial_j\nabla^2\delta$
and
$\partial_i\partial_j\nabla^4\delta$
naturally arise in derivative expansions of tracer bias and in effective field theory descriptions of coarse-grained large-scale structure.
They encode finite-size and curvature-gradient response of tracers and become increasingly relevant toward the nonlinear and ultraviolet regime.

%---------------------------------------------------------
\subsection{From the effective Euler equation to a modified growth equation}

We begin with the standard continuity--Euler--Poisson system in conformal time $\tau$:
\begin{align}
\delta' + \nabla\cdot\left[(1+\delta)\mathbf{v}\right] &= 0, \\
\mathbf{v}' + \mathcal{H}\mathbf{v} + (\mathbf{v}\cdot\nabla)\mathbf{v} &= -\nabla\phi + \mathbf{a}_{\rm eff}, \\
\nabla^2\phi &= \frac{3}{2}\mathcal{H}^2\Omega_m(\tau)\,\delta,
\end{align}
where $\mathcal{H}=a'/a$ is the conformal Hubble rate and a prime denotes $d/d\tau$.

A lowest-order gradient expansion consistent with isotropy allows
\begin{equation}
\mathbf{a}_{\rm eff}
=
-\frac{c_2^2(\tau)}{a^2}\nabla\delta
+
\frac{c_4^2(\tau)}{a^4}\nabla\nabla^2\delta
+\dots,
\label{eq:aeff_full}
\end{equation}
where $c_2$ and $c_4$ are effective coefficients encoding short-range response. { Here the symbols $c_2^2$ and $c_4^2$ denote signed response coefficients; they need not be positive.}

Linearising the system and defining $\theta\equiv\nabla\cdot\mathbf{v}$,
the divergence of the Euler equation yields in configuration space:
\begin{equation}
\theta'(\mathbf{x})+\mathcal{H}\,\theta(\mathbf{x})
=
-\frac{3}{2}\mathcal{H}^2\Omega_m(\tau)\,\delta(\mathbf{x})
-\frac{c_2^2(\tau)}{a^2}\,\nabla^2\delta(\mathbf{x})
+\frac{c_4^2(\tau)}{a^4}\,\nabla^4\delta(\mathbf{x})\,,
\label{eq:euler_div_config}
\end{equation}
where $\theta\equiv\nabla\cdot\mathbf{v}$ and $\nabla^4\equiv(\nabla^2)^2$.

In Fourier space,  the same relation reads:
\begin{equation}
\theta' + \mathcal{H}\theta
=
-\frac{3}{2}\mathcal{H}^2\Omega_m\,\delta
+
\frac{c_2^2}{a^2}k^2\delta
+
\frac{c_4^2}{a^4}k^4\delta.
\end{equation}

Using the linear continuity equation $\delta'=-\theta$,
one obtains the second-order equation
\begin{equation}
\delta'' + \mathcal{H}\delta'
-\frac{3}{2}\mathcal{H}^2\Omega_m\,\delta
+
\frac{c_2^2}{a^2}k^2\delta
+
\frac{c_4^2}{a^4}k^4\delta
=0.
\end{equation}

{ Expressing the evolution in terms of derivatives with respect to $N\equiv\ln a$ (dots), we use
\begin{equation}
\frac{d}{d\tau}=\mathcal{H}\frac{d}{dN},
\qquad
\mathcal{H}=aH,
\end{equation}
so that
\begin{equation}
\delta'=\mathcal{H}\dot{\delta},
\qquad
\delta''=\mathcal{H}^2\ddot{\delta}+\mathcal{H}'\dot{\delta}.
\end{equation}
Since
\begin{equation}
\frac{\mathcal{H}'}{\mathcal{H}^2}=1+\frac{d\ln H}{d\ln a},
\end{equation}
dividing the complete evolution equation by $\mathcal{H}^2=a^2H^2$ gives
\begin{equation}
\ddot{D}
+
\left(2+\frac{d\ln H}{d\ln a}\right)\dot{D}
-
\frac{3}{2}\Omega_m(a)D
+
\frac{c_2^2(a)}{a^4H^2(a)}k^2 D
+
\frac{c_4^2(a)}{a^6H^2(a)}k^4 D
=0.
\label{eq:growth_raw}
\end{equation}}

%---------------------------------------------------------
\subsection{Early-time behaviour and late-time activation}

{ Equation~\eqref{eq:growth_raw} makes clear that constant coefficients $c_2$ and $c_4$ would lead to an unphysical enhancement of the effective gradient corrections at early times. During matter domination, $H^2\propto a^{-3}$, so that the coefficients multiplying the $k^2$ and $k^4$ operators scale respectively as
\begin{equation}
\frac{1}{a^4H^2}\propto a^{-1},
\qquad
\frac{1}{a^6H^2}\propto a^{-3}.
\end{equation}
The higher-derivative term would therefore become increasingly dominant toward early times if $c_4$ were kept constant. This behaviour is not appropriate for the present interpretation, in which these operators describe late-time effective short-range response rather than primordial modifications of the dynamics.}

To enforce the intended interpretation, we introduce a smooth
late-time activation function
\begin{equation}
S(a)
=
\frac{1}{2}
\left[
1+\tanh\left(\frac{\ln(a/a_\star)}{\Delta}\right)
\right],
\label{eq:activation}
\end{equation}
{ which sets a chosen transition epoch and width. It is a phenomenological ansatz, not a consequence of the spectral kernels.}
{ We further absorb the explicit $a^{-4}H^{-2}$ and $a^{-6}H^{-2}$ factors into dimensionless response functions $\beta_2(a)$ and $\beta_4(a)$, yielding the phenomenological growth equation}
\begin{equation}
\ddot{D}
+
\left(2+\frac{d\ln H}{d\ln a}\right)\dot{D}
-
\frac{3}{2}\Omega_m(a)D
+
S(a)
\left[
\beta_2(a)\left(\frac{k}{k_{\rm ref}}\right)^2
+
\beta_4(a)\left(\frac{k}{k_{\rm ref}}\right)^4
\right]
D
=0.
\label{eq:growth_uv_final}
\end{equation}
Here $k_{\rm ref}$ denotes a reference scale, typically chosen near the mesh Nyquist frequency. { This reference choice does not establish the validity of a truncated gradient expansion at the mesh cutoff. The parametrisations agree when $S\beta_2=c_2^2 k_{\rm ref}^2/(a^4H^2)$ and $S\beta_4=c_4^2 k_{\rm ref}^4/(a^6H^2)$.}

\begin{figure}[t]
  \centering
  \includegraphics[width=0.85\textwidth]{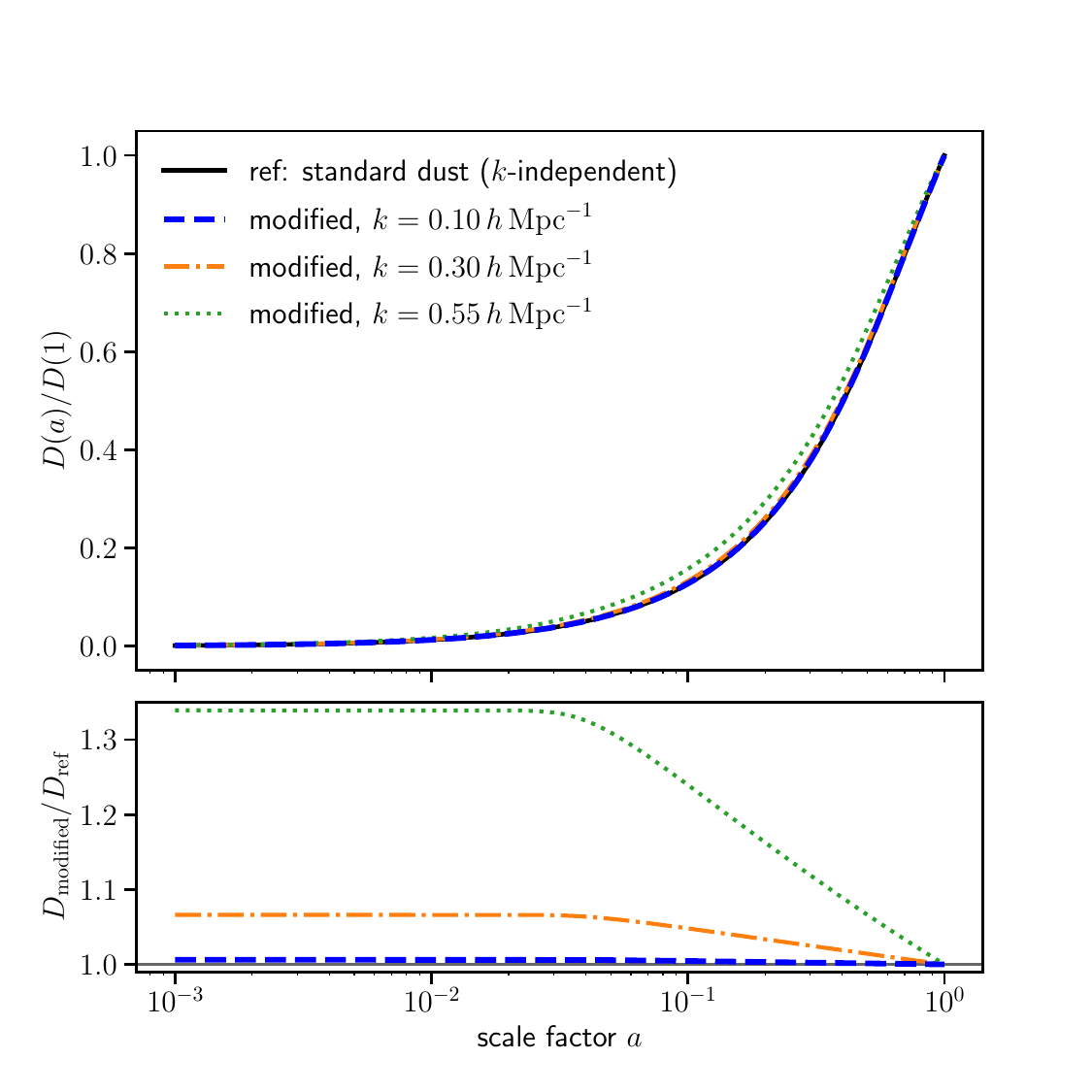}
  \caption{\textbf{Scale-dependent growth with late-time UV activation.}
  {\bf Top panel:} Linear growth factor $D(a)$ obtained by solving Eq.~\eqref{eq:growth_uv_final} for a flat $\Lambda$CDM background.
  The solid black curve shows the standard dust solution (scale independent).
  Dashed coloured curves include scale-dependent UV response terms proportional to $(k/k_{\rm ref})^2$ and $(k/k_{\rm ref})^4$,
  activated at late times through the smooth switching function $S(a)$ defined in Eq.~\eqref{eq:activation}.
  {\bf Bottom panel:} Ratio $D_{\rm modified}/D_{\rm ref}$ highlighting deviations.
  Deviations remain negligible at early times ($a\ll a_\star$), confirming that the dust growing mode is recovered prior to the onset
  of nonlinear clustering. At late times, high-$k$ modes experience controlled deviations, illustrating how such corrections become
  relevant toward the UV limit of coarse-resolution solvers.}
  \label{fig:growth_uv}
\end{figure}

% AUTHOR CHECK: a_star=0.03 corresponds to z~32. Confirm the plotted value and provide the figure parameters.
Figure~\ref{fig:growth_uv} shows numerical solutions of Eq.~\eqref{eq:growth_uv_final} for representative wavenumbers, demonstrating that the UV response remains negligible at early times ($a\ll a_\star\sim3\times10^{-2}$) and becomes increasingly important toward late times, with stronger deviations at larger $k$.

%---------------------------------------------------------
\subsection{Physical interpretation}

Equation~\eqref{eq:growth_uv_final} { provides an illustrative scale-dependent modification} of the growth rate.
For $\beta_n>0$, high-$k$ growth is suppressed,
mimicking effective stress or coarse-grained pressure-like regularisation.
For $\beta_n<0$, high-$k$ growth is enhanced,
which can phenomenologically describe additional short-range collapse
associated with relaxation-type updates.

{ Because $S(a)\rightarrow 0$ at early times, provided that $\beta_2$ and $\beta_4$ remain bounded,}
the evolution asymptotically reduces to the standard dust growth equation
for $a\ll a_\star$.
In numerical solutions, a small constant offset between the modified
and dust growth factors may be observed at early times.
This reflects the chosen normalization of $D(a)$
(e.g.\ enforcing $D(a=1)=1$),
rather than a physical modification of early-time dynamics.
The shape of the growing mode remains consistent with dust evolution
in the limit $a\rightarrow 0$,
{ for growing-mode initial conditions. This early-time limit does not by itself establish stability at high $k$.}

In particular, the early-time solutions remain within the growing-mode subspace of the dust equation, differing only by an overall multiplicative factor that is absorbed by late-time normalization.

%=========================================================
%=========================================================

%=========================================================
\section{Cosmic web classification and its relation to nonlocal bias}
\label{sec:nonlocal_bias_link}

A central motivation for this work is that cosmic web classification is not merely a morphological descriptor:
it encodes the same tensor invariants that appear in long-range nonlocal bias expansions.
This insight was developed explicitly in hierarchical cosmic web assembly-bias approaches
\citep{Kitaura_2022,Coloma_2024}.
Here we summarize and generalize that connection, and show how the spectral hierarchy provides a systematic extension.

This section is a conceptual interpretation that motivates the UV levels; the hierarchy itself is defined purely by the spectral operators.

\subsection{Tidal tensor invariants and long-range nonlocal bias}

Consider the tidal tensor $\mathcal{T}_{ij}=\partial_i\partial_j\phi$ and its eigenvalues $\lambda_1\ge\lambda_2\ge\lambda_3$.
Define the invariants
\begin{equation}
\label{eq:inv}
I_1=\lambda_1+\lambda_2+\lambda_3,\qquad
I_2=\lambda_1\lambda_2+\lambda_1\lambda_3+\lambda_2\lambda_3,\qquad
I_3=\lambda_1\lambda_2\lambda_3.
\end{equation}
For $\nabla^2\phi=\delta$, one has $I_1=\delta$.
Writing the traceless tidal tensor
\begin{equation}
s_{ij}\equiv \mathcal{T}_{ij}-\frac{1}{3}\delta_{ij}^{K}\delta,
\end{equation}
the scalar nonlocal operators in bias expansions can be expressed in terms of invariants, e.g.
\begin{equation}
\label{eq:deltas_invariants}
\delta = I_1,\qquad
s^2\equiv s_{ij}s_{ji}=\frac{2}{3}I_1^2-2I_2,\qquad
s^3\equiv s_{ij}s_{jk}s_{ki}=-I_1I_2+3I_3+\frac{2}{9}I_1^3.
\end{equation}
Thus, any classification that segments space by the signs and relative magnitudes of $(\lambda_1,\lambda_2,\lambda_3)$
implicitly conditions on $(I_1,I_2,I_3)$ and therefore on $(\delta,s^2,s^3)$.
This explains why the $i=-2$ level can encode long-range nonlocal bias information
without explicitly inserting nonlocal operators into a bias model \citep{Kitaura_2022,Coloma_2024}.

\subsection{Short-range nonlocality and curvature / density-derivative tensors}

Short-range nonlocal bias contributions can be organized via density-derivative tensors,
e.g. $\Gamma_{ij}\equiv \partial_i\partial_j\delta$.
Define the traceless part
\begin{equation}
\mu_{ij}\equiv \Gamma_{ij}-\frac{1}{3}\delta_{ij}^{K}\nabla^2\delta,
\end{equation}
and invariants $\mu^2\equiv \mu_{ij}\mu_{ji}$, $\mu^3\equiv \mu_{ij}\mu_{jk}\mu_{ki}$.
In hierarchical assembly-bias approaches one can sub-classify $i=-2$ regions using curvature information,
thereby conditioning on short-range operator content \citep{Coloma_2024}.

Within the spectral hierarchy, $i=0$ is the natural curvature (density Hessian) level, and $i=2,4$ extend this idea to higher derivatives:
these levels systematically correspond to higher-derivative operator families in bias expansions,
offering a controlled ladder of short-range information aligned with renormalised operator content.

\subsection{Why hierarchy helps: positive definite bias modelling via catalogue subdivision}

A practical issue with truncated explicit bias expansions is that they can yield negative predicted densities or introduce many parameters. {In addition, modelling the bias as a local parametric formula implies assuming the same galaxy formation physics everywhere in the Universe, and therefore the need of introducing complex models to account for different regimes, such as positive/negative threshold bias to boost/suppress galaxy formation.}
The hierarchical cosmic web approach addresses these issues by {sub}dividing a tracer catalogue into subsets based on web environment
(e.g. $4\times4=16$ regions in a two-level scheme { combining the tidal tensor ($i=-2$) and the density Hessian ($i=0$)}), and applying a {local} parametric model in each subset. { The subdivision therefore incorporates part of the nonlocal environmental dependence into the conditioning variables, allowing the local relation to be calibrated separately in each region \citep{McDonald09,Coloma_2024}. Positivity requires a non-negative conditional count model.}
Each subset behaves like a separate tracer population, analogously to mass binning, but now organized by physically meaningful nonlocal information
\citep{Kitaura_2022,Coloma_2024}. {In this way, the local bias model employed in each environment typically simplifies significantly, naturally accounting for the phenomena driving galaxy formation in each cosmic web environment}.
The spectral hierarchy generalizes the set of possible conditioning variables in a way that remains directly interpretable in bias-operator language.

%=========================================================
\section{Simulation Data}
\label{sec:data}

To investigate the spectral hierarchical cosmic web across a broad range of spatial scales,
we analyse two complementary simulation datasets.
A high-resolution particle-mesh simulation is used to explore the morphological
structure of the hierarchy in the matter field, while a large-volume $N$-body simulation
provides halo catalogues suitable for statistical tests of the information content
relevant for large-scale structure analyses and mock generation.

\subsection{High-resolution particle-mesh simulation}
\label{sec:pm_bench}

We first analyse a particle-mesh simulation designed to resolve the detailed morphology
of the cosmic web hierarchy.
 The simulation volume has side length $L=100\,h^{-1}\mathrm{Mpc}$ 
and contains $400^3$ particles, corresponding to a spatial resolution $\mathrm{d}L = 0.25\,h^{-1}\mathrm{Mpc}$.

The particle-mesh evolution is performed using the code
\textsc{FastPM} \citep{2016MNRAS.463.2273F}.
Initial conditions are generated using second-order Lagrangian perturbation theory
(2LPT) at redshift $z=99$ and evolved to $z=1$ using $50$ time steps.
The force resolution is chosen to match the particle spacing.
 The density field is analysed on a cubic mesh with cell size $\Delta x = 0.25\,h^{-1}\mathrm{Mpc}$. The corresponding Nyquist frequency is $k_{\rm Nyquist} \equiv (\pi/\Delta x)
\simeq 12.5\,h\,\mathrm{Mpc}^{-1}$.

We construct the matter overdensity field $\delta(\mathbf{x})$
using a CIC mass-assignment scheme with tetrahedral tessellation
\citep{Abel_2012}. All cosmic-web classification fields are evaluated
on the same mesh.

This high-resolution setup allows us to investigate the detailed
morphological behaviour of the spectral hierarchy in the matter field.

\subsection{\textsc{Abacus} dark matter and halo catalogues}
\label{sec:abacus_data}

To quantify the statistical information content of the hierarchy
for biased tracers, we additionally analyse halo catalogues
from the publicly available \textsc{AbacusSummit} simulation suite
\citep{Maksimova_2021}, produced with the \textsc{Abacus} $N$-body code
\citep{Garrison_2018,Garrison2021}.

The \textsc{AbacusSummit} simulations consist of 97 cosmological models
centred around a base cosmology consistent with Planck 2018
\citep{Planck2018}. Each simulation evolves $6912^3$ particles in a
volume of $(2\,h^{-1}\mathrm{Gpc})^3$, corresponding to a particle mass
of approximately $m_p \simeq 2\times10^{9}\,h^{-1}M_\odot$,  sufficient to resolve DESI-like tracers \citep{Levi2019}.

In this work we analyse one realisation of the base cosmology. 
We focus on z=1.1 for the information-content tests; the companion RLPT paper uses both z=0.2 and z=1.1 for gravity benchmarks. 
We use both the dark matter density field
and the corresponding halo catalogue.

For the halo-based tests we select the most massive haloes until
reaching a target number density of $n_h = 3\times10^{-3}\,(h\,\mathrm{Mpc}^{-1})^{3}$, 
yielding approximately $8\times10^{6}$ haloes in the full simulation box.
% AUTHOR CHECK: n_h=0.003 and L=2000 imply 24 million haloes; 8 million imply n_h=0.001. Confirm the actual catalogue before changing either value.
 For the field-level analysis we work on a coarse cubic mesh with cell
size $\Delta x \simeq 5.5\,h^{-1}\mathrm{Mpc}$,
which matches the resolution regime typically used in fast
large-volume forward modelling and mock production. The corresponding Nyquist frequency is $k_{\rm Nyquist} \equiv (\pi/\Delta x)
\simeq 0.57\,h\,\mathrm{Mpc}^{-1}$.

The halo overdensity field $\delta_h(\mathbf{x})$ is constructed using
a nearest-grid-point assignment scheme.
The matter field used for comparison is obtained by down-sampling the
white-noise field of the corresponding \textsc{Abacus} simulation to a
mesh of $360^3$ cells and evolving it with ALPT
\citep{Kitaura_2013}. Cosmic-web classification fields are evaluated on
the same mesh. 
{The motivation for using ALPT in this second, low-resolution case is to investigate a practical and scalable setup in which the cross-correlation between the different cosmic-web regions is computed from a fast approximate gravity solver and compared directly to accurate tracer fields, here represented by the \textsc{Abacus} halo catalogue, rather than relying on a down-sampled version of the expensive \textsc{Abacus} dark matter field itself.
This provides a realistic test of the regime relevant for mock production and field-level inference, where one typically has access to approximate dark matter realizations but wishes to recover the environmental information needed to model high-fidelity tracers.
It also connects directly to the companion paper on ridged LPT, where we study subgrid modelling as a practical route to reconstruct the high-resolution halo distribution from low-resolution approximate gravity solvers. }

%=========================================================
\section{Method: web classification, information-theoretic compression, and cross-correlation}
\label{sec:method}

In this section we describe how the spectral hierarchy is implemented in practice
and how its information content is quantified.
Our goal is to assess whether the categorical environmental information encoded
at each hierarchy level $i$ retains predictive power for halo clustering
across Fourier scales relevant for fast mock production.

The procedure consists of three steps:
(i) constructing generalized web classifications from the hierarchy tensors
$\mathcal{H}^{(i)}_{ab}$,
(ii) compressing the categorical web information into a four-value field that
preserves the ordered morphology (void--sheet--filament--knot),
and (iii) measuring the cross-correlation between this compressed web field
and the halo overdensity field in Fourier space.

This pipeline allows us to test, in a controlled and information-theoretic way,
whether the spectral hierarchy captures nontrivial halo-relevant structure
from large scales down to the mesh Nyquist frequency.

\subsection{Generalized eigenvalue web classification at level $i$}

For each hierarchy level $i$, we compute $\delta^{(i)}$ via Eq.~\eqref{eq:kernel_def},
construct $\Hess^{(i)}_{ab}$ via Eq.~\eqref{eq:general_hessian}, and compute its ordered eigenvalues
$\lambda_1^{(i)}$, $\lambda_2^{(i)}$, $\lambda_3^{(i)}$.
{
We define web regions according to the number of eigenvalues exceeding a threshold $\lambda_{\rm th}^{(i)}$.
Assuming the ordering $\lambda_1^{(i)} \ge \lambda_2^{(i)} \ge \lambda_3^{(i)}$, the classification reads
\begin{itemize}
\item \textbf{knots:} $\lambda_3^{(i)} > \lambda_{\rm th}^{(i)}$,
\item \textbf{filaments:} $\lambda_2^{(i)} > \lambda_{\rm th}^{(i)}$ and $\lambda_3^{(i)} \le \lambda_{\rm th}^{(i)}$,
\item \textbf{sheets:} $\lambda_1^{(i)} > \lambda_{\rm th}^{(i)}$ and $\lambda_2^{(i)} \le \lambda_{\rm th}^{(i)}$,
\item \textbf{voids:} $\lambda_1^{(i)} \le \lambda_{\rm th}^{(i)}$\,,
\end{itemize}
\citep[see][]{Zeldovich_1970,Bond_1996nat,Hahn_2007, Forero_2009},
By construction, these four classes are mutually exclusive and collectively exhaustive, so that every cell is assigned to exactly one web type.
}
For the information-content demonstration below we focus on near-zero thresholds, with mild tuning only where needed to stabilize noise at high-derivative levels (as is standard in practical web classifiers).

{
To test whether the web hierarchy carries information relevant to halos across scales, we compress the web classification into a four-value field.
Following hierarchical cosmic web work \citep{Coloma_2024}, we assign the ordered labels
\begin{equation}
w(\mathbf{x})=
\begin{cases}
4,& \text{knots},\\
3,& \text{filaments},\\
2,& \text{sheets},\\
1,& \text{voids},
\end{cases}
\end{equation}
and define a corresponding web density contrast
\begin{equation}
\delta_{\rm web}^{(i)}(\mathbf{x})\equiv \frac{w^{(i)}(\mathbf{x})}{\avg{w^{(i)}}}-1.
\label{eq:web_contrast}
\end{equation}
For Gaussian random fields, the joint distribution of the eigenvalues of the deformation tensor derived by Doroshkevich 1970 \citep{1970Afz.....6..581D} implies that regions with one and two positive eigenvalues (sheet- and filament-like configurations) dominate the volume, while fully compressing or expanding configurations (knots and voids) are much less probable.
The equidistant spacing adopted here should therefore not be interpreted as reflecting equal volume fractions, but rather as the simplest ordered encoding of the four web classes.
This choice avoids introducing additional assumptions or optimization into the compression, and is sufficient for the present purpose of demonstrating that even a minimal categorical representation retains substantial cross-correlation with the tracer field.
More general or optimized labelings could be explored in future work if required for specific practical applications.}
%=========================================================

{ For two fields $A$ and $B$, we measure the Fourier-space correlation coefficient
\begin{equation}
C_{AB}(k)=\frac{\operatorname{Re}\langle\hat A^*(\mathbf q)\hat B(\mathbf q)\rangle_k}
{\sqrt{\langle|\hat A(\mathbf q)|^2\rangle_k\langle|\hat B(\mathbf q)|^2\rangle_k}}
=\frac{P_{AB}(k)}{\sqrt{P_{AA}(k)P_{BB}(k)}}.
\label{eq:pearson_fourier}
\end{equation}
Here $\langle\cdot\rangle_k$ denotes an average over Fourier modes in a shell centred on $k$. It differs from the spatial mean $\langle w\rangle$ used to define the web contrast. We use $A=\delta_{\rm tr}$ and $B=\delta_{\rm web}^{(i)}$.}

\section{Results: evidence for information in the hierarchy}
\label{sec:results}

In this section we investigate the behaviour of the spectral cosmic-web
hierarchy in both simulations introduced above.

We first examine the qualitative morphology of the hierarchy using the
high-resolution \textsc{FastPM} simulation.
This allows us to visualise how different hierarchy levels probe
distinct spatial scales and geometric features of the matter field.

We then quantify the statistical information content of the hierarchy
using the large-volume \textsc{Abacus} simulation by measuring
cross-correlations between the web-compressed fields and the halo
density field.

\subsection{Visual impression of hierarchical information}

\begin{figure}
    \vspace{-1.2cm}
    \centering
    \begin{tabular}{cc}
\subfigure{\includegraphics[width=0.5\textwidth]{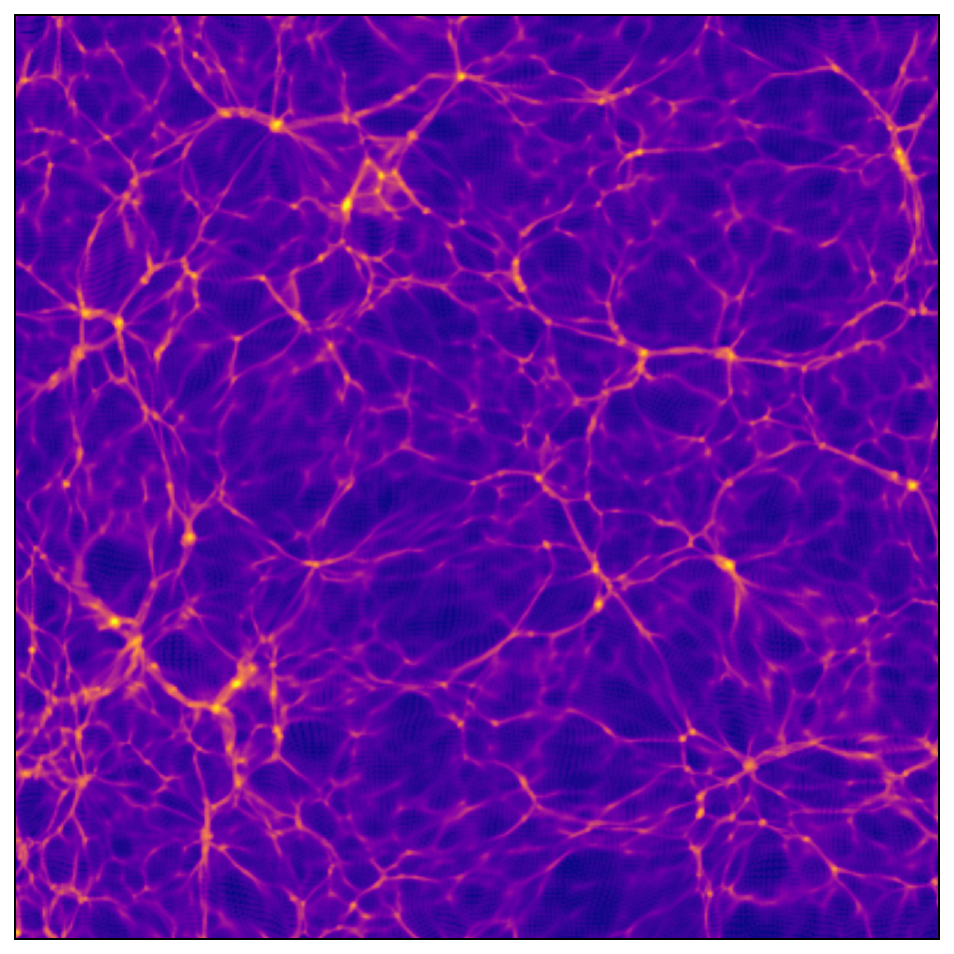}}
\put(-218,207){\fcolorbox{white}{white}{\color{black}$\delta_{\rm DM}-\mathrm{FastPM}$}}
\put(-218,192){\fcolorbox{white}{white}{\color{black}$z=1$}}
\hspace{-0.4cm}
&\subfigure{\includegraphics[width=0.5\textwidth]{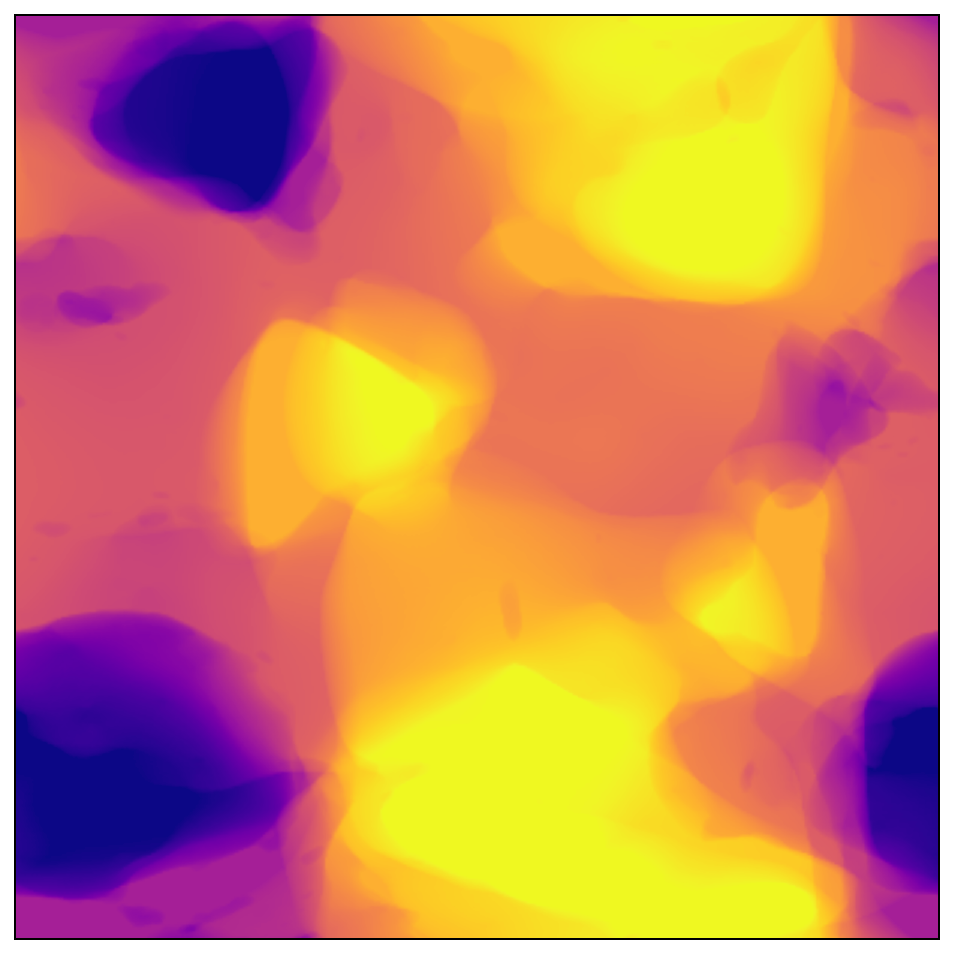}}
\put(-218,207){\fcolorbox{white}{white}{\color{black}$-k^{-4}$}}
\vspace{-0.4cm}
\\
\subfigure{\includegraphics[width=0.5\textwidth]{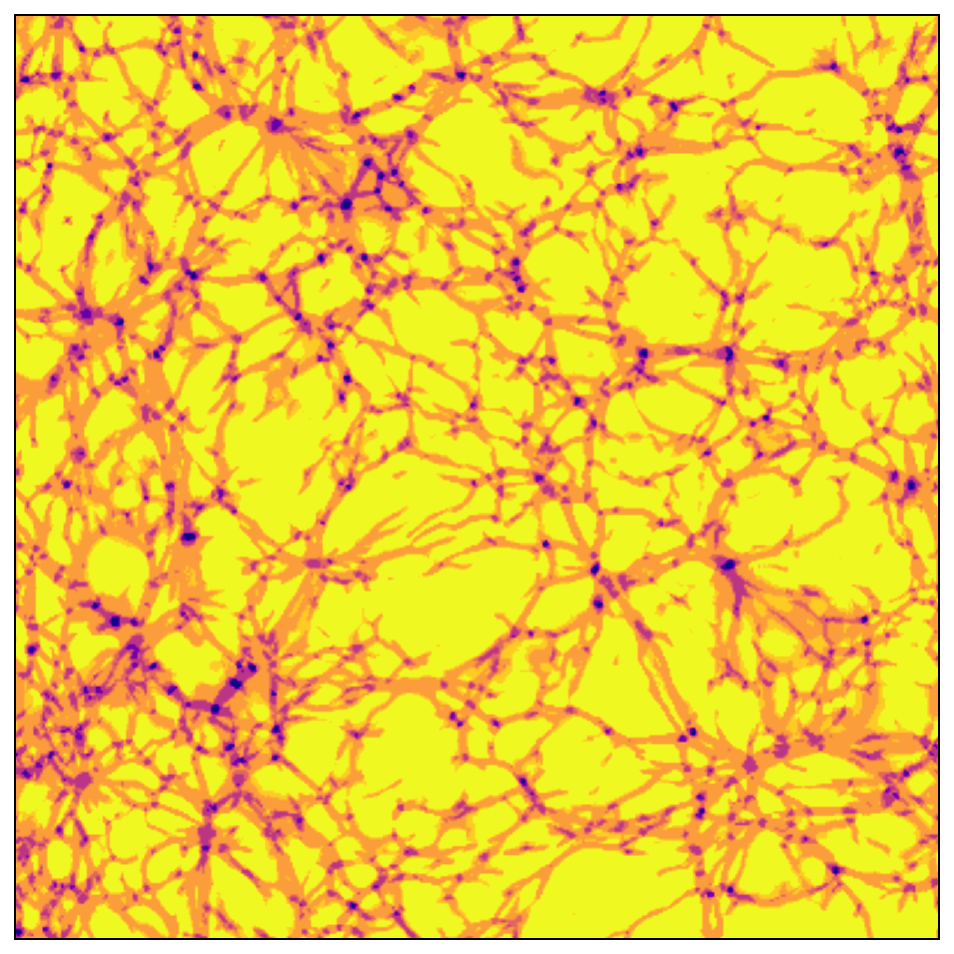}}  \put(-218,207){\fcolorbox{white}{white}{\color{black}$-k^{-2}$}}\hspace{-0.4cm} &
\subfigure{\includegraphics[width=0.5\textwidth]{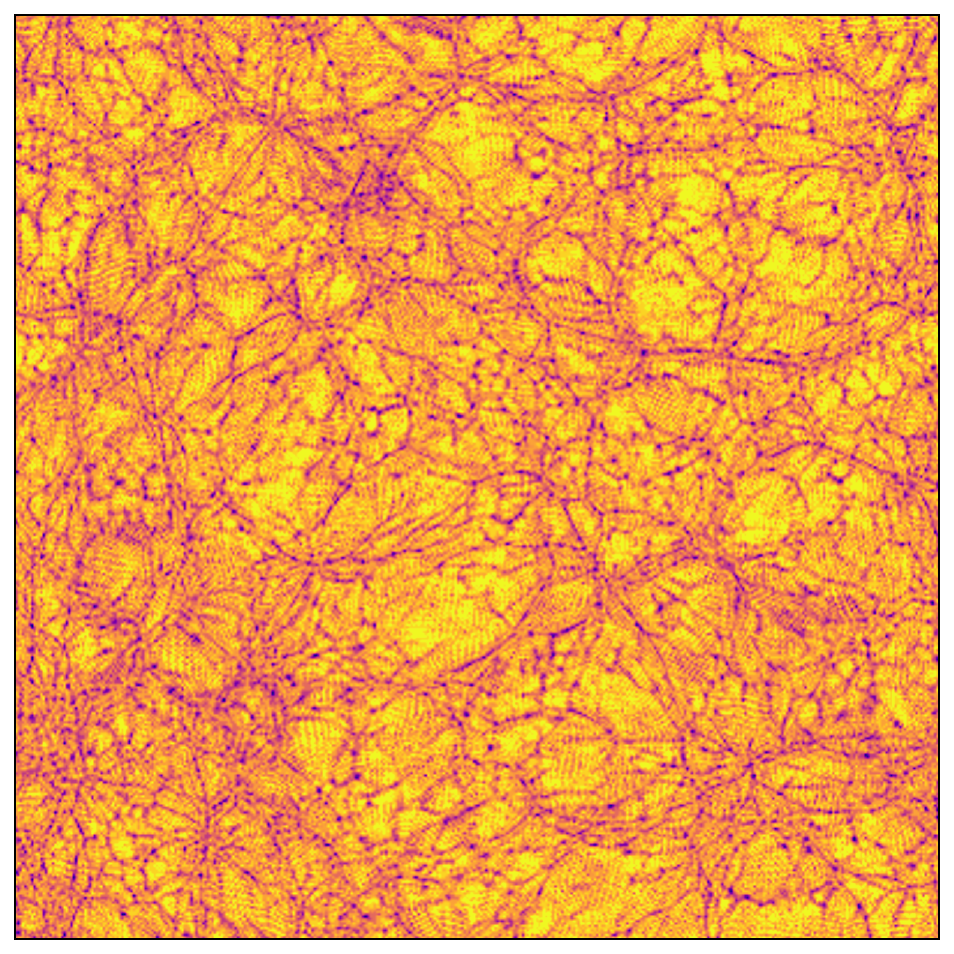}}\put(-218,207){\fcolorbox{white}{white}{\color{black}$-k^{0}$}}
\vspace{-0.4cm}
\\
\subfigure{\includegraphics[width=0.5\textwidth]{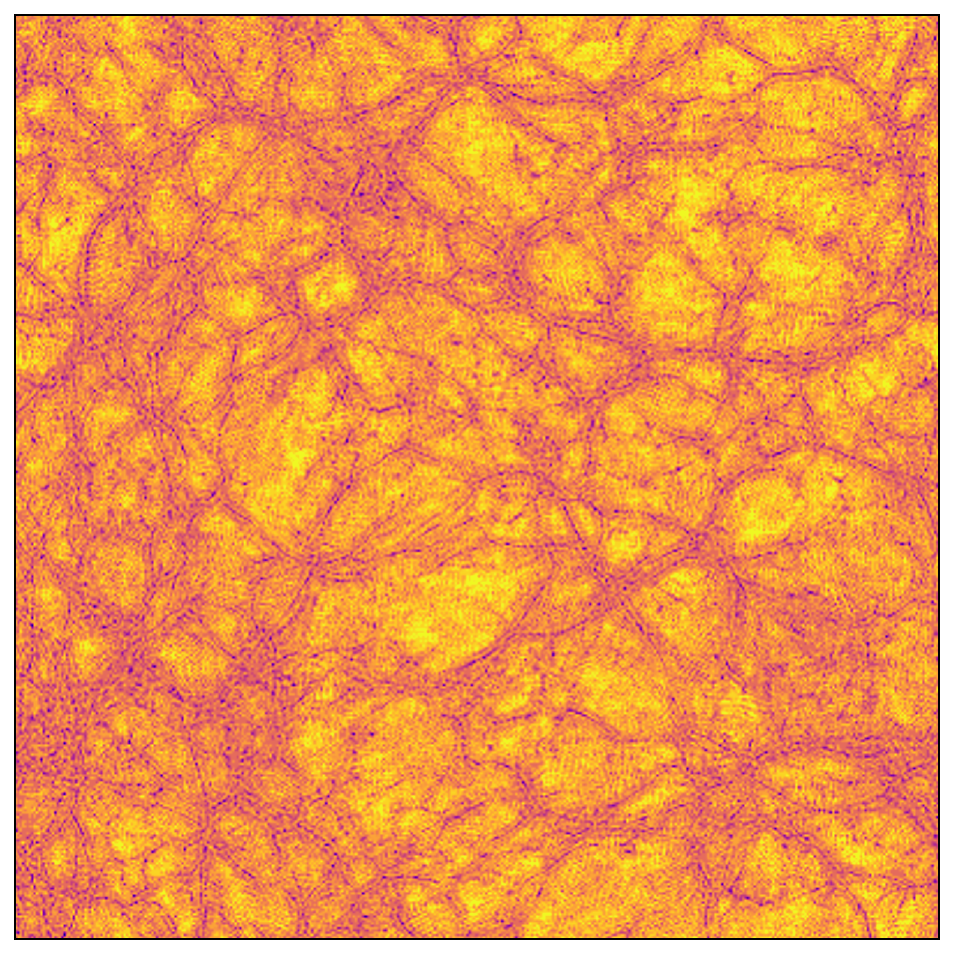}} \put(-218,207){\fcolorbox{white}{white}{\color{black}$-k^{2}$}} \hspace{-0.4cm} &
\subfigure{\includegraphics[width=0.5\textwidth]{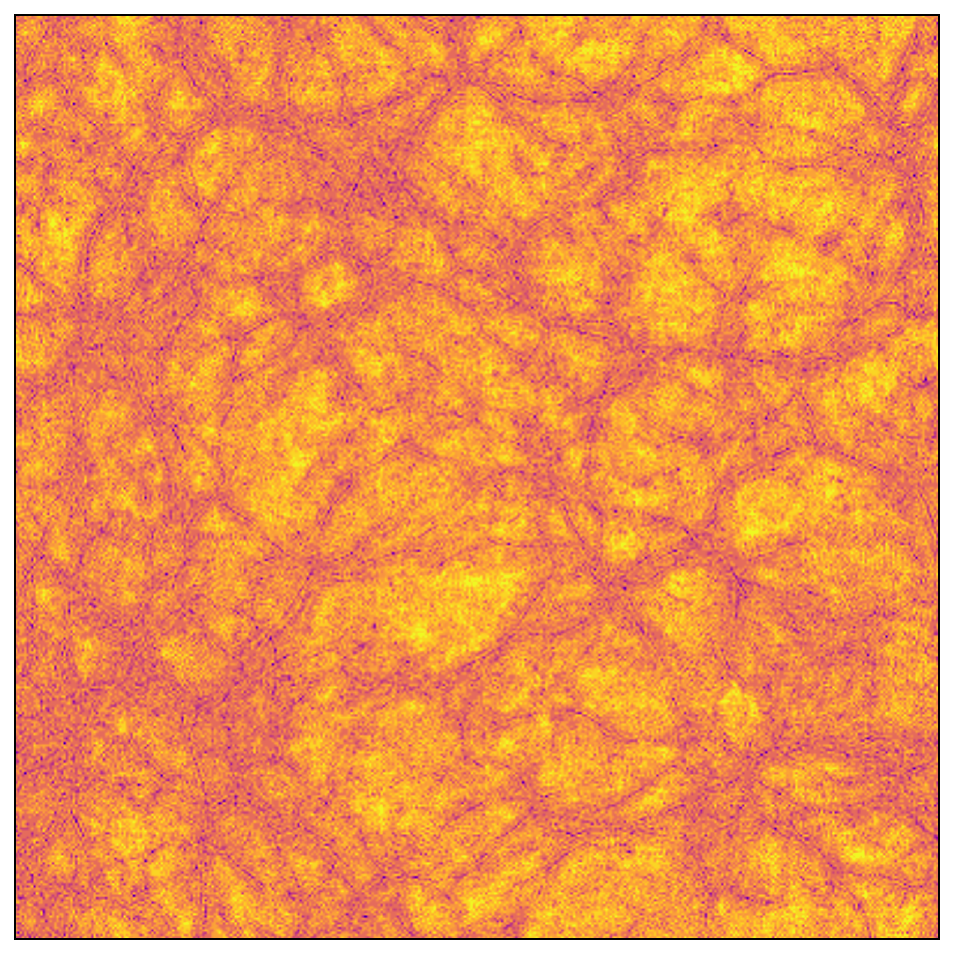}}\put(-218,207){\fcolorbox{white}{white}{\color{black}$-k^{4}$}}
\vspace{-0.4cm}
\end{tabular}
\put(-455,311){\rotatebox[]{90}{\large$\longrightarrow$}}    \put(-460,220){\rotatebox[]{90}{\large${\rm L}=100\,h^{-1}\,{\rm Mpc}$}}
\put(-455,115){\rotatebox[]{90}{\large$\longleftarrow$}}
\put(-400,328){\rotatebox[]{0}{\large$\longrightarrow$}}    \put(-380,328){\rotatebox[]{0}{\large${\rm d}L=0.25\,h^{-1}\,{\rm Mpc}$}}
\put(-283,328){\rotatebox[]{0}{\large$\longleftarrow$}}
\caption{{\bf Top left}: \textsc{FastPM}  at $z=1$ assigned to a mesh. The side length is $100\,h^{-1}{\rm Mpc}$ and the cell size is ${\rm d}L\simeq 0.25\,h^{-1}{\rm Mpc}$.
Other panels show the cosmic web classification contrast corresponding to the kernel hierarchy: ({\bf top right}) $-k^{-4}$,
({\bf middle left}) $-k^{-2}$, ({\bf middle right}) $-k^0$, ({\bf bottom left}) $-k^{2}$, ({\bf bottom right}) $-k^{4}$.}
\label{fig:cosmic-web-fastpm}
\end{figure}

Figure~\ref{fig:cosmic-web-fastpm} illustrates the behaviour of the
spectral hierarchy in the high-resolution \textsc{FastPM} simulation.

The top left panel shows the underlying matter density field.
The subsequent panels display cosmic web classifications obtained from
different hierarchy levels.
The top right panel corresponds to the $i=-4$ classification
($-k^{-4}$ kernel), which strongly weighs long-wavelength modes.
As expected, this level highlights only very broad spatial modulations
of the density field.
Fine filamentary structure is largely suppressed, indicating that
$i=-4$ primarily probes ultra-large-scale morphology.

The mid left panel shows the classical potential-based
classification ($i=-2$), corresponding to the standard tidal-tensor
cosmic web.
This level recovers the familiar network of filaments and knots
associated with anisotropic gravitational collapse.
For the high-resolution case we apply a logarithmic transformation of
the density field prior to the web classification in order to reduce
the impact of strongly nonlinear density peaks.

The subsequent panels---mid right ($i=0$), bottom left ($i=2$),
and bottom right ($i=4$)---progressively highlight smaller-scale
structure.
At $i=0$, curvature information sharpens ridge and peak morphology
relative to the potential web.
The $i=2$ and $i=4$ levels further enhance small-scale curvature,
producing increasingly localized ridge and knot patterns.

This progression demonstrates the ladder structure of the hierarchy:
as the kernel index $i$ increases, sensitivity shifts systematically
from large-scale tidal morphology to increasingly local curvature
features of the density field.

\subsection{Cross power spectrum analysis}

To quantify the information carried by the different hierarchy levels,
we compute the cross power spectrum between the compressed web fields
$\delta_{\rm web}^{(i)}$ and the underlying tracer field.

\subsubsection{High-resolution matter field.}

We first analyse the high-resolution \textsc{FastPM} simulation,
where the web hierarchy is evaluated directly on the matter density
field at resolution $\Delta x=0.25\,h^{-1}\mathrm{Mpc}$.
The cross-correlation coefficient between the matter overdensity
$\delta$ and the web-compressed fields $\delta_{\rm web}^{(i)}$
reveals a clear scale-dependent hierarchy of information content.

At intermediate nonlinear scales around $k \sim 1\,h\,\mathrm{Mpc}^{-1}$,
all hierarchy levels remain highly correlated with the matter field.
The measured correlations are
$C^{(-2)} \simeq 0.87$,
$C^{(0)} \simeq 0.91$,
$C^{(2)} \simeq 0.92$,
and $C^{(4)} \simeq 0.87$.
At smaller scales the hierarchy becomes increasingly important.
For example, at $k \sim 10\,h\,\mathrm{Mpc}^{-1}$ the correlations are
$C^{(-2)} \simeq 0.14$,
$C^{(0)} \simeq 0.53$,
$C^{(2)} \simeq 0.68$,
and $C^{(4)} \simeq 0.58$.

Thus, while the classical tidal-tensor level ($i=-2$) rapidly loses
predictive power at high $k$, the higher hierarchy levels retain
substantial correlation with the matter field.
Toward the ultraviolet end of the spectrum, and especially close to the mesh Nyquist frequency,
the $i=4$ level provides the strongest correlation,
indicating that the highest-derivative hierarchy level captures the most localized morphological information in the density field.
This behaviour demonstrates that higher levels of the spectral
hierarchy capture small-scale curvature information that is absent in
the traditional tidal classification.

\subsubsection{Halo field on coarse meshes.}

We perform an analogous analysis using the halo catalogue from the
\textsc{AbacusSummit} simulation.

Although \textsc{AbacusSummit} is intrinsically a significantly higher-resolution
$N$-body simulation than the \textsc{FastPM} run used above,
the halo field is intentionally analysed on a much coarser mesh in the present study.
This choice reflects the resolution regime typical of fast forward models
and mock-generation pipelines.

In this setup the higher hierarchy levels already dominate the
information content at moderately nonlinear scales.
The transition occurs around $k \gtrsim 0.27\,h\,\mathrm{Mpc}^{-1}$.
At $k \simeq 0.5\,h\,\mathrm{Mpc}^{-1}$, the correlations with the halo field are $C^{(-2)} \simeq 0.22,\quad
C^{(0)} \simeq 0.52,\quad
C^{(2)} \simeq 0.64,\quad
C^{(4)} \simeq 0.64$.

Figure~\ref{fig:ck_dm_web} summarizes the central quantitative result of this work.
In the high-resolution \textsc{FastPM} matter field (top panel), the higher hierarchy levels
$i=0,2,4$ overtake the classical tidal-web level $i=-2$ at $k\gtrsim0.6\,h\,\mathrm{Mpc}^{-1}$,
showing that the additional derivative levels encode substantially more information in the nonlinear regime.
At $k\simeq1\,h\,\mathrm{Mpc}^{-1}$ the correlations are already slightly higher for $i=0$ and $i=2$ than for $i=-2$,
while at $k\simeq10\,h\,\mathrm{Mpc}^{-1}$ the gain becomes dramatic:
$C^{(-2)}\simeq0.14$, compared with $C^{(0)}\simeq0.53$, $C^{(2)}\simeq0.68$, and $C^{(4)}\simeq0.58$.
A similar trend is found in the \textsc{Abacus} halo analysis on coarse meshes (bottom panel),
where levels higher than $i=-2$ begin to dominate at $k\gtrsim0.27\,h\,\mathrm{Mpc}^{-1}$.
At $k\simeq0.5\,h\,\mathrm{Mpc}^{-1}$ the classical tidal-web level yields only
$C^{(-2)}\simeq0.22$, whereas the higher levels reach
$C^{(0)}\simeq0.52$ and $C^{(2)}\simeq C^{(4)}\simeq0.64$.
This demonstrates that higher levels of the spectral hierarchy provide a major gain in information content toward nonlinear scales.

Together these results demonstrate that the spectral hierarchy
systematically extends the range of scales over which cosmic-web
information remains relevant.
While the classical tidal web captures large-scale anisotropic
collapse, the higher derivative levels encode additional small-scale
structure that becomes increasingly important toward nonlinear
regimes.

{
Although correlations between hierarchy levels are expected because
their tensors derive from the same underlying field, this does not
make their web classifications redundant for tracer modelling.
Their different spectral weights probe distinct aspects of the
environment, and the classification at one level does not uniquely
determine that at another. Quantitative evidence for this
complementarity was provided by Coloma-Nadal et al. \citep[][]{Coloma_2024}: supplementing
the potential-based web ($i=-2$) with the density-curvature web
($i=0$) improved the simultaneous reproduction of the halo PDF,
power spectrum and reduced bispectrum, with the latter not used
in the calibration. The present results extend this investigation
to higher hierarchy levels. Separating the contributions of
statistical correlations between levels from those associated
with nonlinear gravitational evolution remains an interesting
question beyond the scope of this work.
}

\subsection{PDF characterization and tracer footprint}

\begin{figure}[ht!]
\vspace{-0.6cm}
    \centering
    \includegraphics[width=\textwidth]{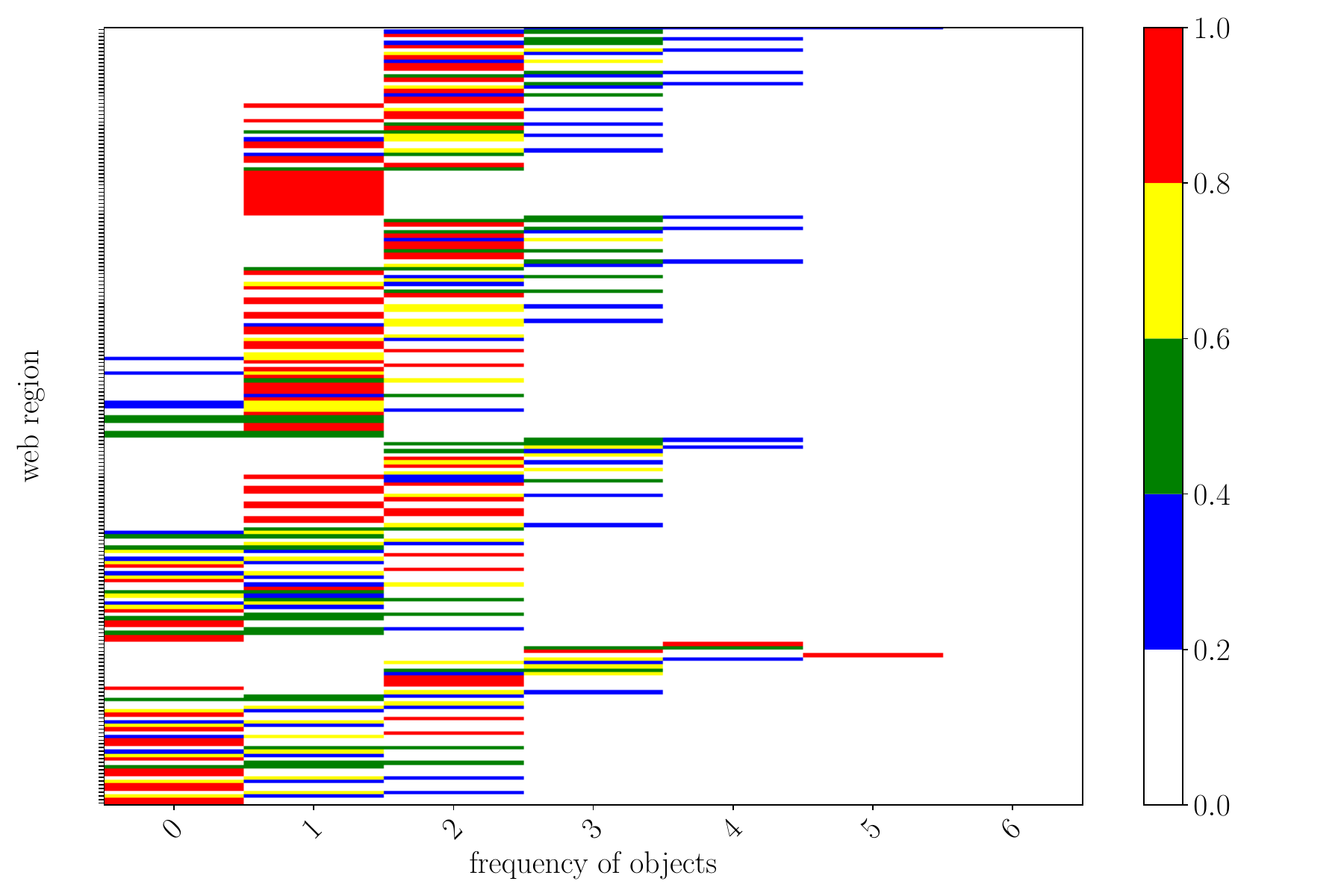}
    \put(-415,280){$0$}
    \put(-425,26){$255$}
\caption{
Environmental footprint of the halo population across the spectral hierarchy.
The figure shows the probability distribution functions (PDFs) of halo occupation
within the 256 cosmic web regions obtained from four hierarchy levels
($i=-2, 0, 2, 4$), each subdivided into four morphological classes
(voids, sheets, filaments, knots).
For clarity, the PDF values are binned in intervals of width $0.2$ in the range $[0,1]$.
Each bin corresponds to a distinct combination of web classifications across levels,
forming a discrete multi-scale environmental signature of the tracer population.
The non-uniform occupation across regions demonstrates that halos respond differently
to long-range tidal structure, curvature, and higher-derivative local environment.
This multi-level PDF can therefore be interpreted as the environmental
{footprint} of the tracer population and provides a compact basis
for bias characterization and environmental modelling.
}
    \label{fig:web_frequency}
\end{figure}

A complementary characterization of the halo distribution is provided
by the probability distribution functions (PDFs) in the different cosmic web regions,
shown in Fig.~\ref{fig:web_frequency}.
For simplicity and visual clarity,
we restrict the binning to intervals of width $0.2$ in the range $[0,1]$.
Considering four hierarchy levels from $i=-2$ to $i=4$,
each with four morphological classes,
one obtains $4^4=256$ distinct cosmic web regions.

The resulting PDFs can be interpreted as the {environmental footprint}
of the tracer population.
Each tracer type populates the 256 regions with a characteristic distribution,
encoding its response to long-range tidal structure, curvature,
and higher-derivative local environment.
Rather than expanding in a large number of explicit nonlocal operators,
one may describe tracers through their empirical footprint across hierarchy levels.

In this sense, the spectral hierarchy provides a compact and physically interpretable
basis for tracer characterization.
It compresses multi-scale environmental information into a discrete,
low-dimensional structure that nevertheless retains substantial correlation
with the halo density field across Fourier scales. 

\begin{figure}
    \vspace{-2cm}    
    \centering
    \begin{tabular}{cc}
\subfigure{\includegraphics[width=0.5\textwidth]{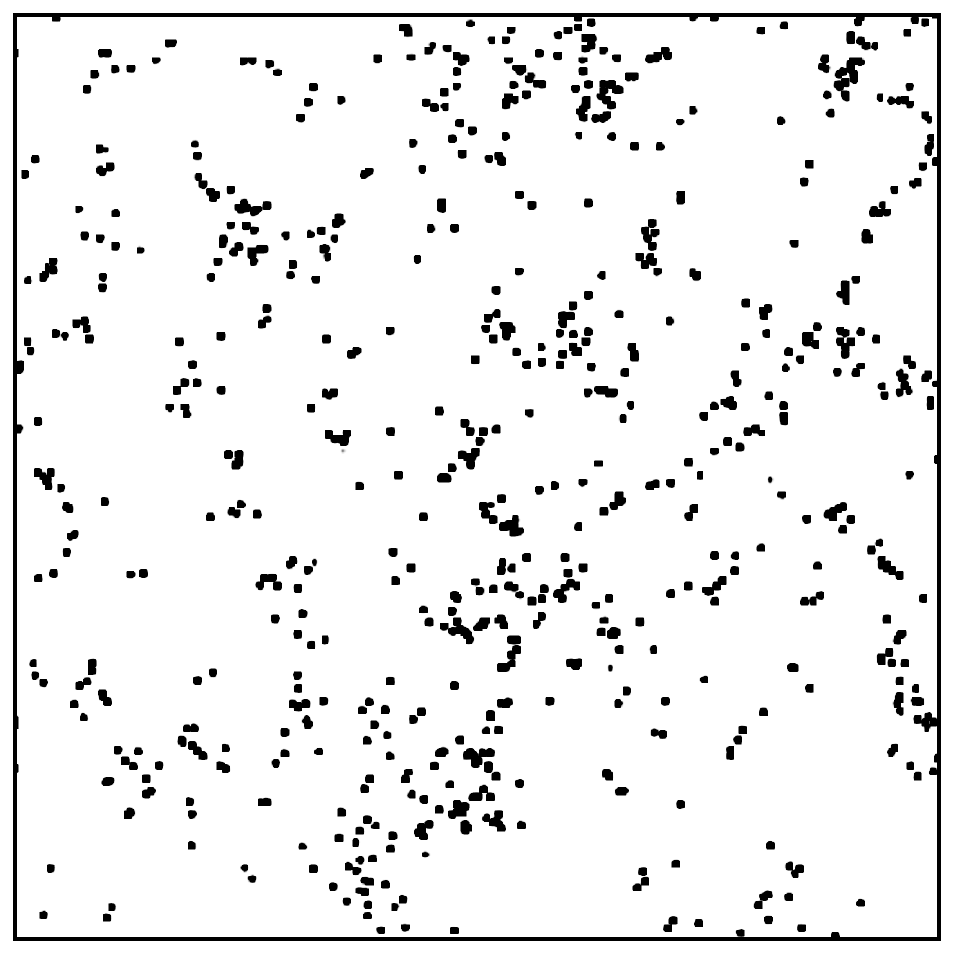}} \put(-218,207){\fcolorbox{white}{white}{\color{black}h$-$ABACUS}} 
\hspace{-0.4cm} & 
\subfigure{\includegraphics[width=0.5\textwidth]{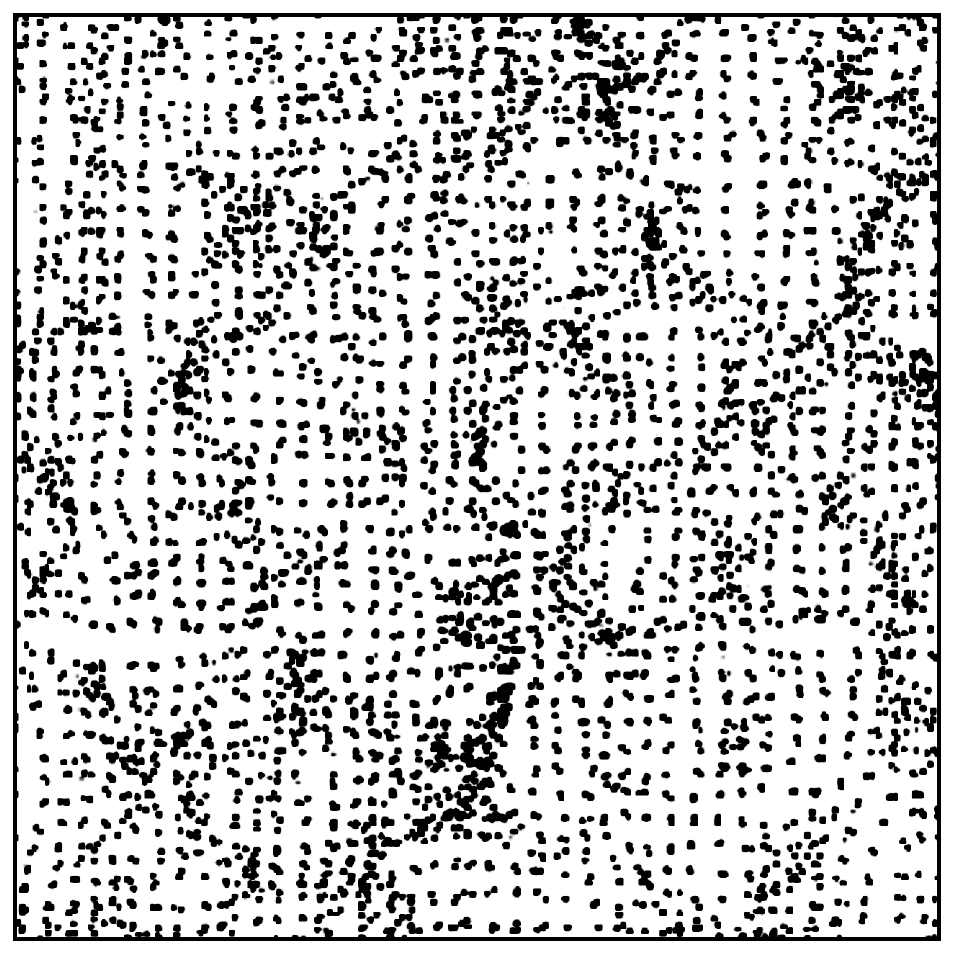}}\put(-218,207){\fcolorbox{white}{white}{\color{black}${\rm DM-ALPT}$}}\vspace{-0.3cm}\\
\subfigure{\includegraphics[width=0.5\textwidth]{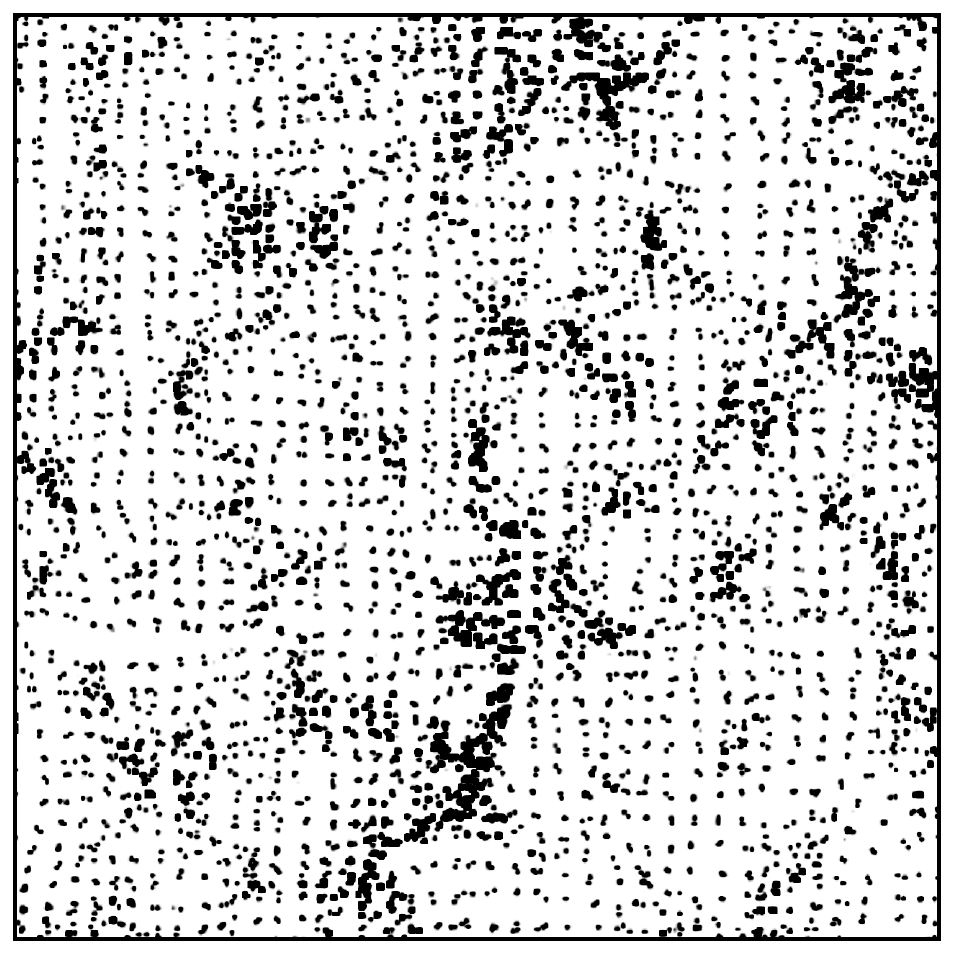}}\put(-218,207){\fcolorbox{white}{white}{\color{black}${\rm DM-ALPT}\, (-k^{-2})$}}
\hspace{-0.4cm} & 
\subfigure{\includegraphics[width=0.5\textwidth]{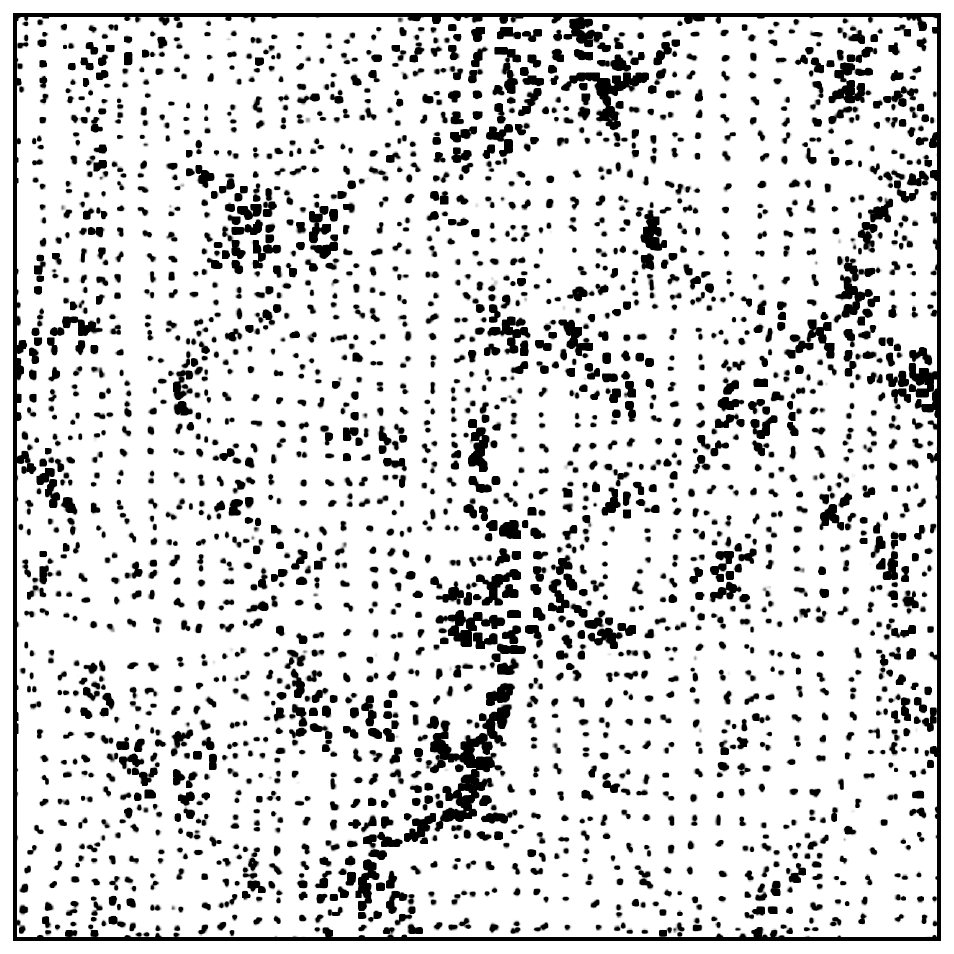}}\put(-218,207){\fcolorbox{white}{white}{\color{black}${\rm DM-ALPT}\, (-k^{-2}, -k^{0})$}}
\vspace{-0.3cm}\\
\subfigure{\includegraphics[width=0.5\textwidth]{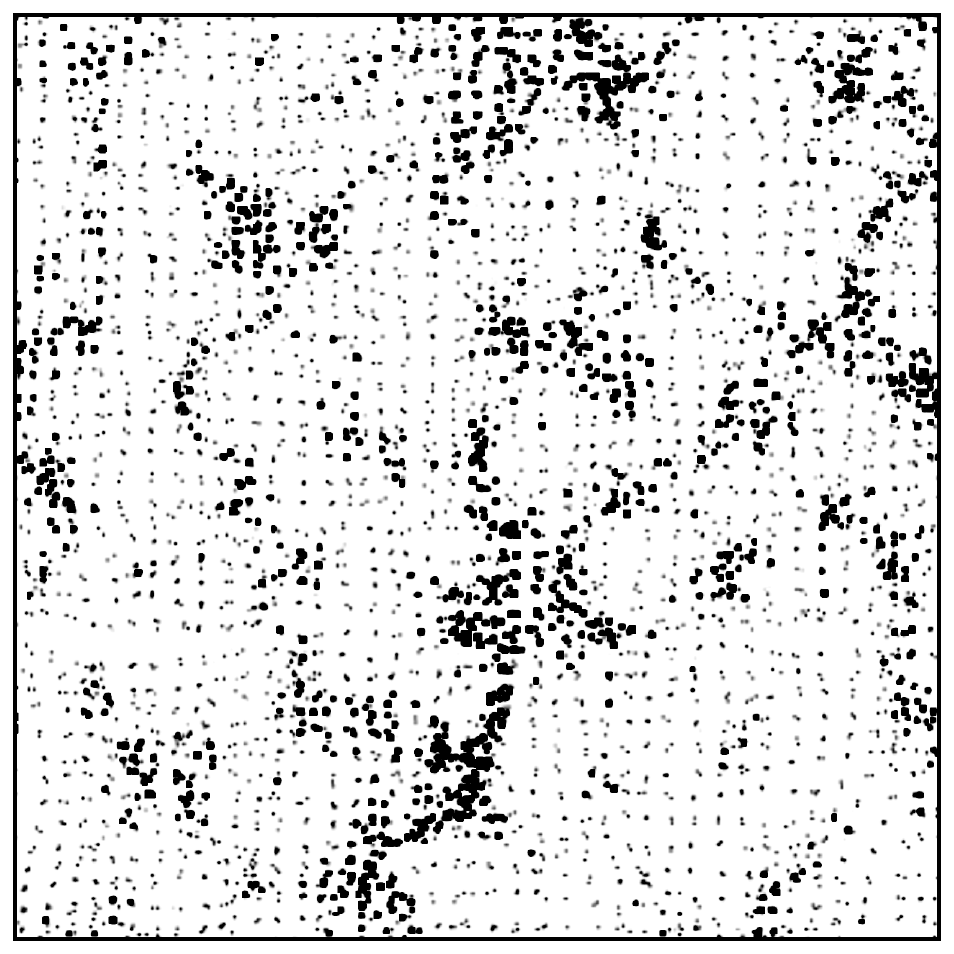}}\put(-218,207){\fcolorbox{white}{white}{\color{black}${\rm DM-ALPT}\, (-k^{-2}, -k^{0}, -k^2)$}}
\hspace{-0.4cm} & 
\subfigure{\includegraphics[width=0.5\textwidth]{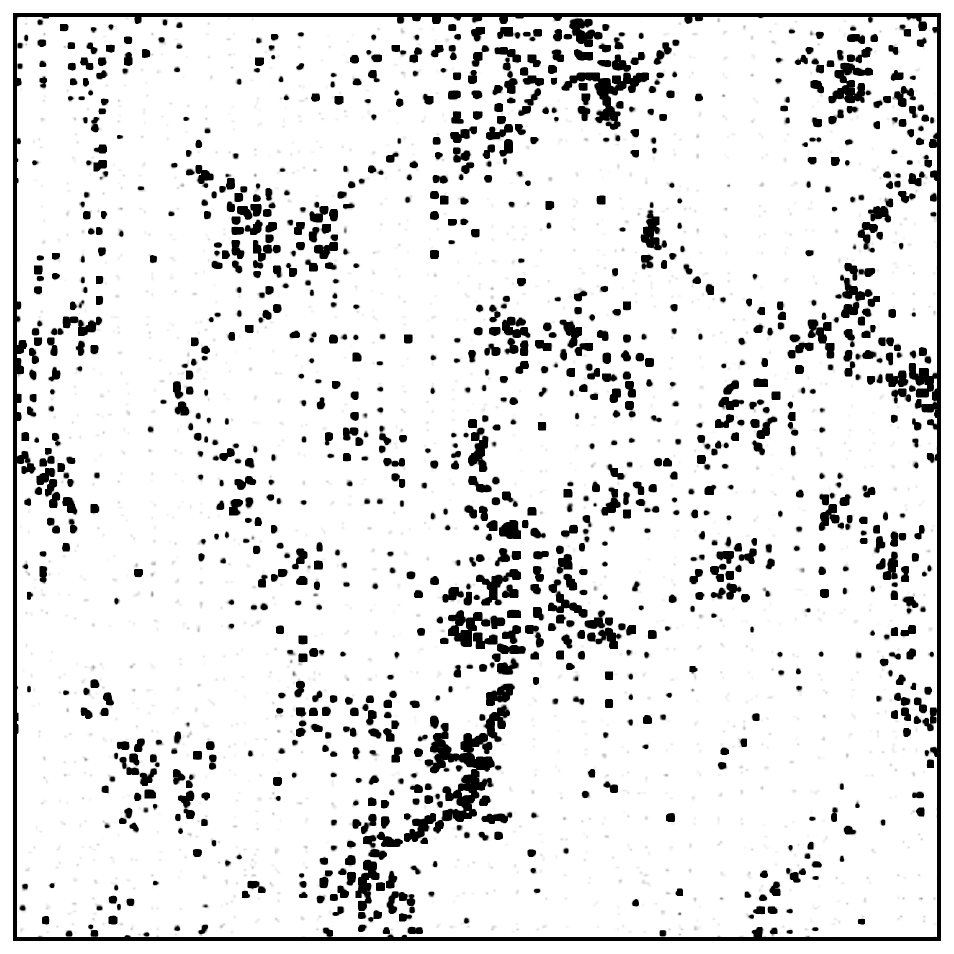}}\put(-218,207){\fcolorbox{white}{white}{\color{black}${\rm DM-ALPT}\, (-k^{-2}, -k^{0}, -k^2, -k^4)$}}
\end{tabular}
\put(-455,317){\rotatebox[]{90}{\large$\longrightarrow$}}    \put(-460,230){\rotatebox[]{90}{\large${\rm L}=250\,[\subset2000]\,h^{-1}\,{\rm Mpc}$}}
\put(-455,125){\rotatebox[]{90}{\large$\longleftarrow$}}
\put(-420,335){\rotatebox[]{0}{\large$\longrightarrow$}}    \put(-400,335){\rotatebox[]{0}{\large${\rm dL}=0.69\,[\subset5.55]\,h^{-1}\,{\rm Mpc}$}}
\put(-260,335){\rotatebox[]{0}{\large$\longleftarrow$}}
\vspace{-0.5cm}
\caption{Slice of a $250\,h^{-1}\mathrm{Mpc}$ subvolume extracted from the \textsc{Abacus}  simulation, represented on a $360^3$ mesh with cell size $d\mathrm{L}=0.69,h^{-1}\mathrm{Mpc}$. The {\bf upper panels} show the \textsc{Abacus} halo number counts ({\bf left}), and the ALPT dark matter particle positions on the high resolution mesh   ({\bf right}). The {\bf remaining panels}  display the dark matter particles after assigning masses according to halo number densities measured in the \textsc{Abacus}  simulation for different cosmic-web regions defined by the spectral hierarchy computed from the ALPT field (see text for details).}
\label{fig:cwnd}
\end{figure}

\begin{figure}[ht!]
\vspace{-1cm}
    \centering
    \begin{tabular}{c}
    \includegraphics[scale=0.5]{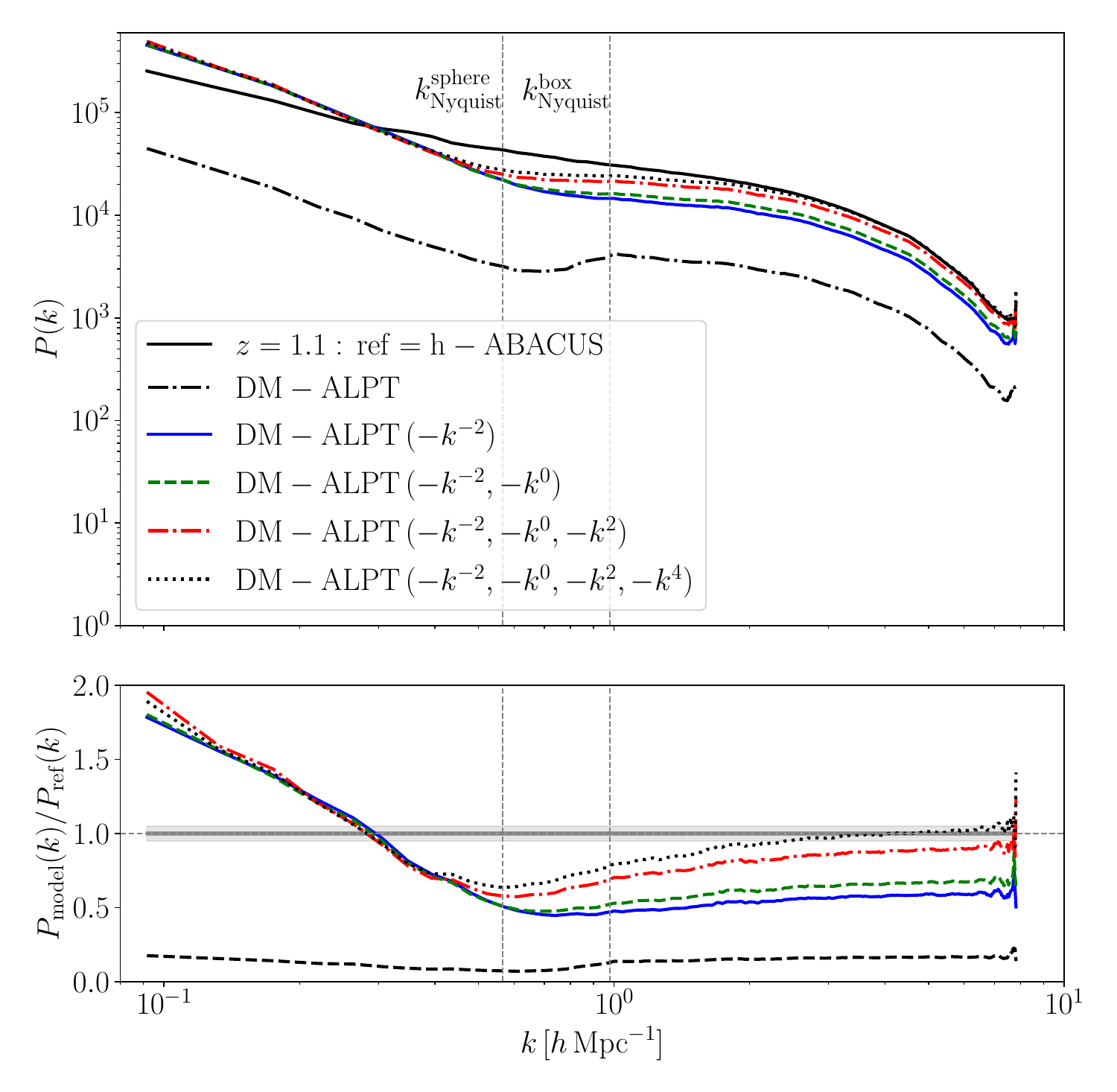}
    \end{tabular}
\caption{Power spectra of the different halo and dark matter fields. The solid line shows the halo distribution from the \textsc{Abacus}  simulation on a mesh with resolution $d\mathrm{L}=0.69,h^{-1}\mathrm{Mpc}$. The dash–dotted line corresponds to the ALPT dark matter particle distribution from the low-resolution mesh gridded onto the high-resolution mesh. The subsequent DM–ALPT curves represent the same dark matter field constrained to match the \textsc{Abacus}  halo number densities in progressively finer cosmic-web hierarchies. 
%We computed the spherical Nyquist frequency as $k_{\rm Nyquist}^{\rm sphere}\equiv \frac{\pi}{\mathrm{d}L}\simeq \frac{\pi}{5.55}\simeq 0.57\,h\,\mathrm{Mpc}^{-1}$, and 
The box Nyquist frequency is $k_{\rm Nyquist}^{\rm box}=k_{\rm Nyquist}^{\rm sphere}\times\sqrt{3}\simeq0.99$.}
    \label{fig:pkhighk}
\end{figure}

In Figs.~\ref{fig:cwnd} and \ref{fig:pkhighk} we illustrate how progressively more small-scale clustering information can be recovered by weighting the ALPT dark matter particles according to the halo number densities measured in the \textsc{Abacus} simulation within increasingly refined cosmic-web partitions, namely 4, 16, 64, and 256 regions.
Figure~\ref{fig:cwnd} shows this visually in a $250\,h^{-1}\mathrm{Mpc}$ subvolume: as the spectral hierarchy is refined, the weighted ALPT particle distribution increasingly resembles the spatial footprint of the halo field.
The same trend is quantified in Fig.~\ref{fig:pkhighk}, where the high-$k$ power spectrum converges toward the \textsc{Abacus} halo clustering to within a few percent as more web regions are included.
By contrast, the discrepancies at low $k$ remain substantial, since in this exercise we only reweight the dark matter particles using halo number densities in web environments, without yet modelling the tracer number counts themselves.
Those large-scale amplitudes would need to be supplied by an effective field-level bias model, whereas the present test isolates the gain in subgrid and small-scale information provided by the spectral cosmic-web hierarchy \citep[for an extension of this exercise see the companion paper]{KitauraSinigaglia_RLPT_2026}.   

\subsection{Implications for fast mock generation}

The key practical implication is that the scale range over which the hierarchy
retains information coincides with the resolution regime typical of
large-volume fast lightcone simulations and mock pipelines.
The spectral hierarchy therefore offers a viable environmental conditioning
basis for fast mock generation:
it captures information that would otherwise require
either finer mesh resolution or more expensive dynamical modelling.

More broadly, this result supports the idea that cosmic web conditioning
can serve as an efficient proxy for nonlocal bias operators,
providing a computationally economical yet physically motivated
route to incorporate environment-dependent effects in forward modelling
and galaxy mock production.
An application of this can be found in the subgrid modelling  \citep[see][]{Forero_2024,KitauraSinigaglia_RLPT_2026}.

%=========================================================

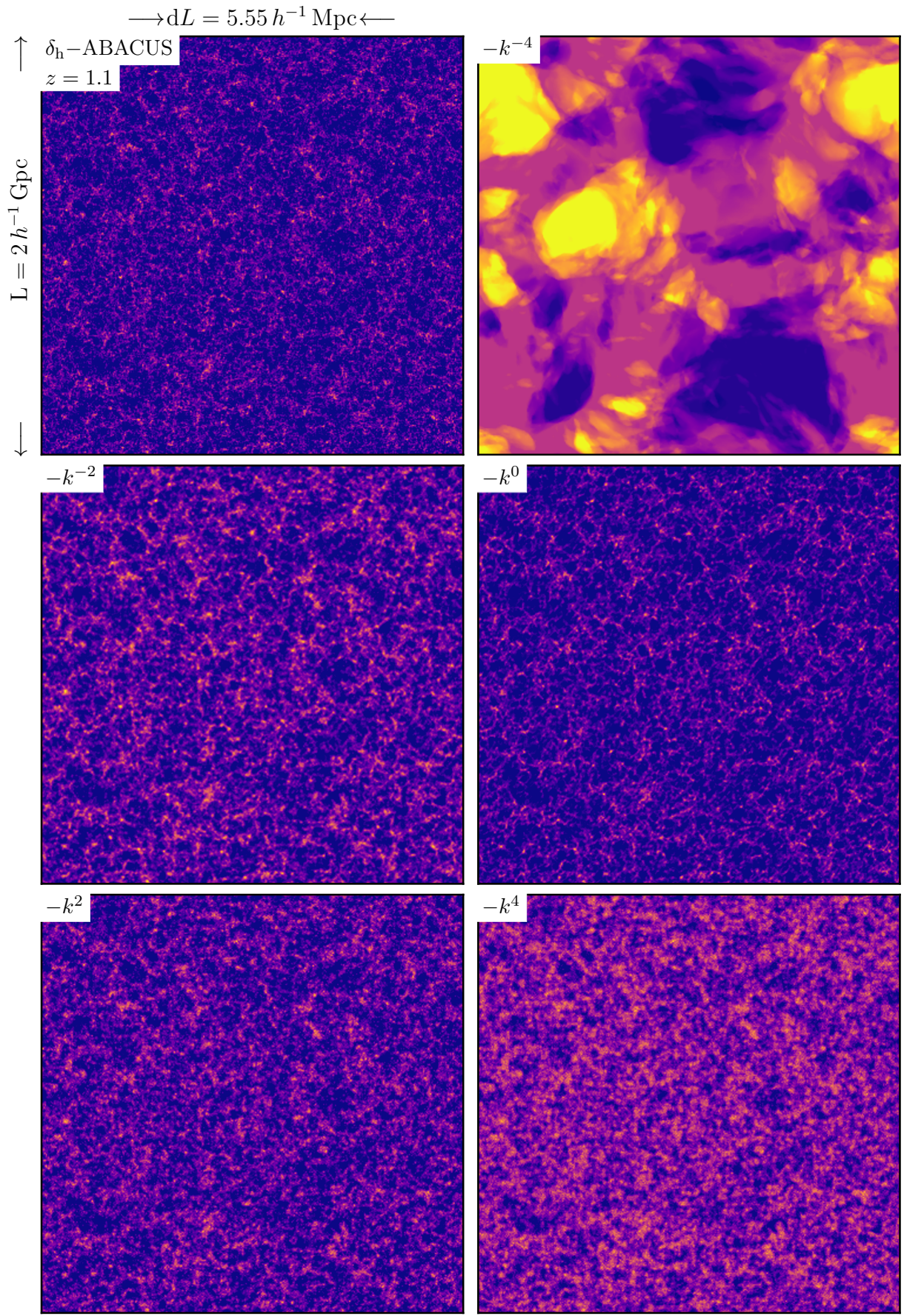
\begin{figure}
    \vspace{-1.2cm}
    \centering
    \begin{tabular}{cc}
\subfigure{\includegraphics[width=0.5\textwidth]{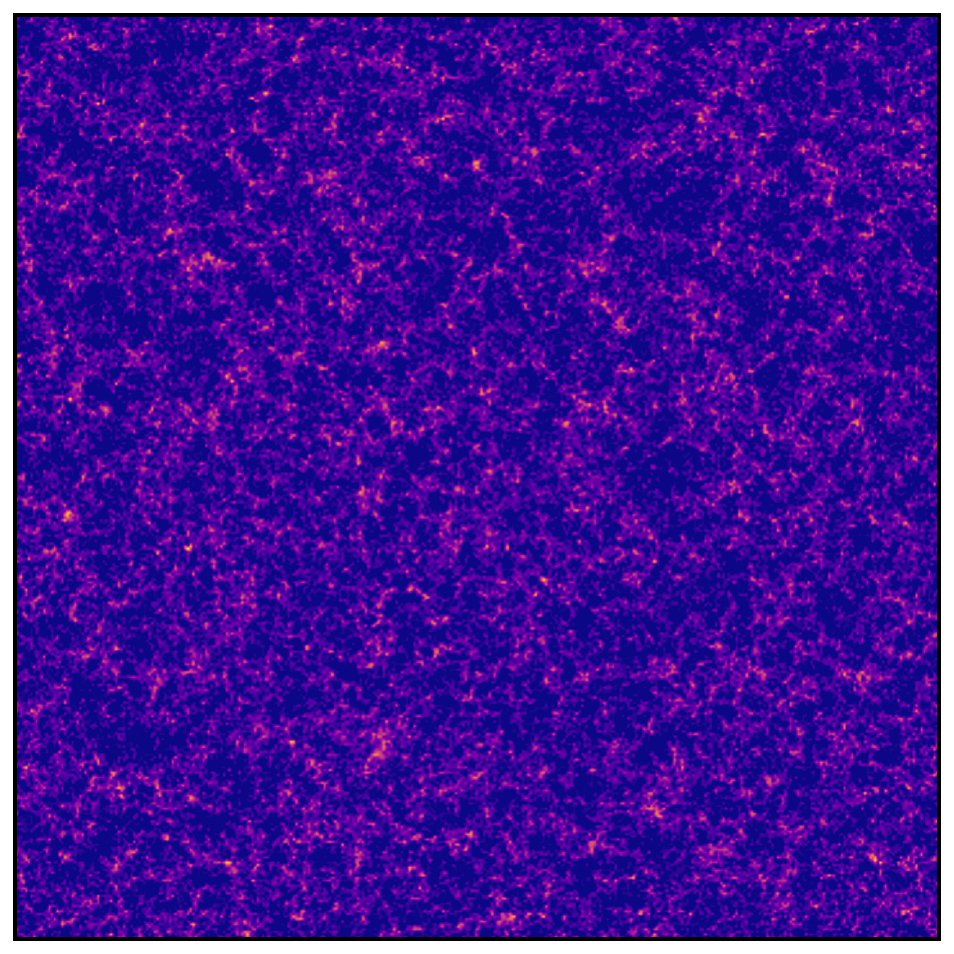}}
\put(-218,207){\fcolorbox{white}{white}{\color{black}$\delta_{\rm h}-$ABACUS}}
\put(-218,192){\fcolorbox{white}{white}{\color{black}$z=1.1$}}
\hspace{-0.4cm}
&\subfigure{\includegraphics[width=0.5\textwidth]{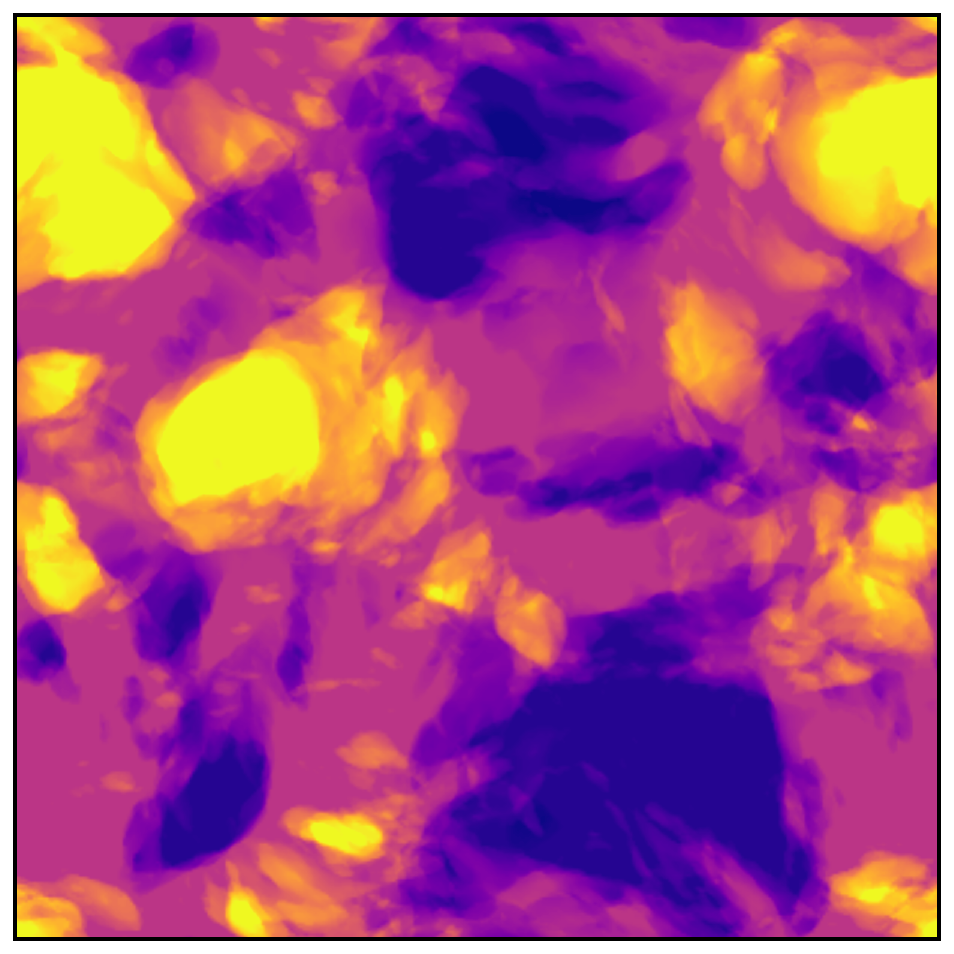}}\put(-218,207){\fcolorbox{white}{white}{\color{black}$-k^{-4}$}}
\vspace{-0.4cm}
\\
\subfigure{\includegraphics[width=0.5\textwidth]{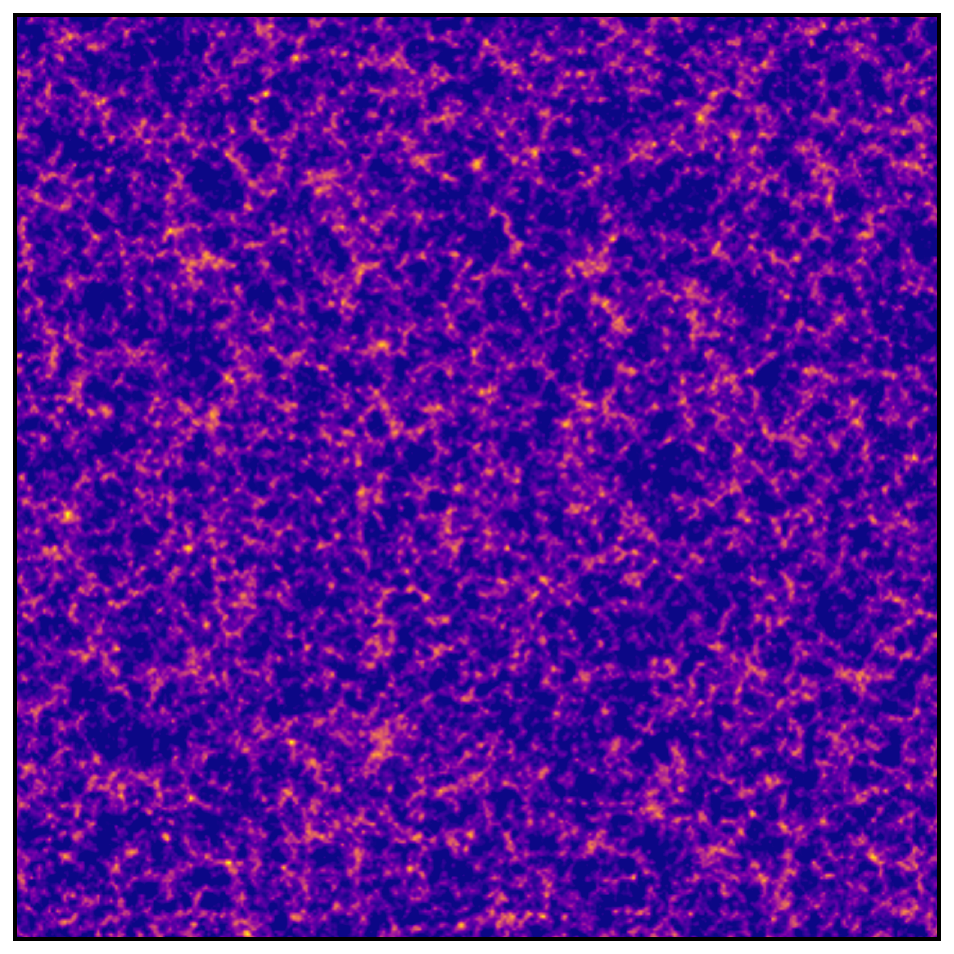}}  \put(-218,207){\fcolorbox{white}{white}{\color{black}$-k^{-2}$}}\hspace{-0.4cm} &
\subfigure{\includegraphics[width=0.5\textwidth]{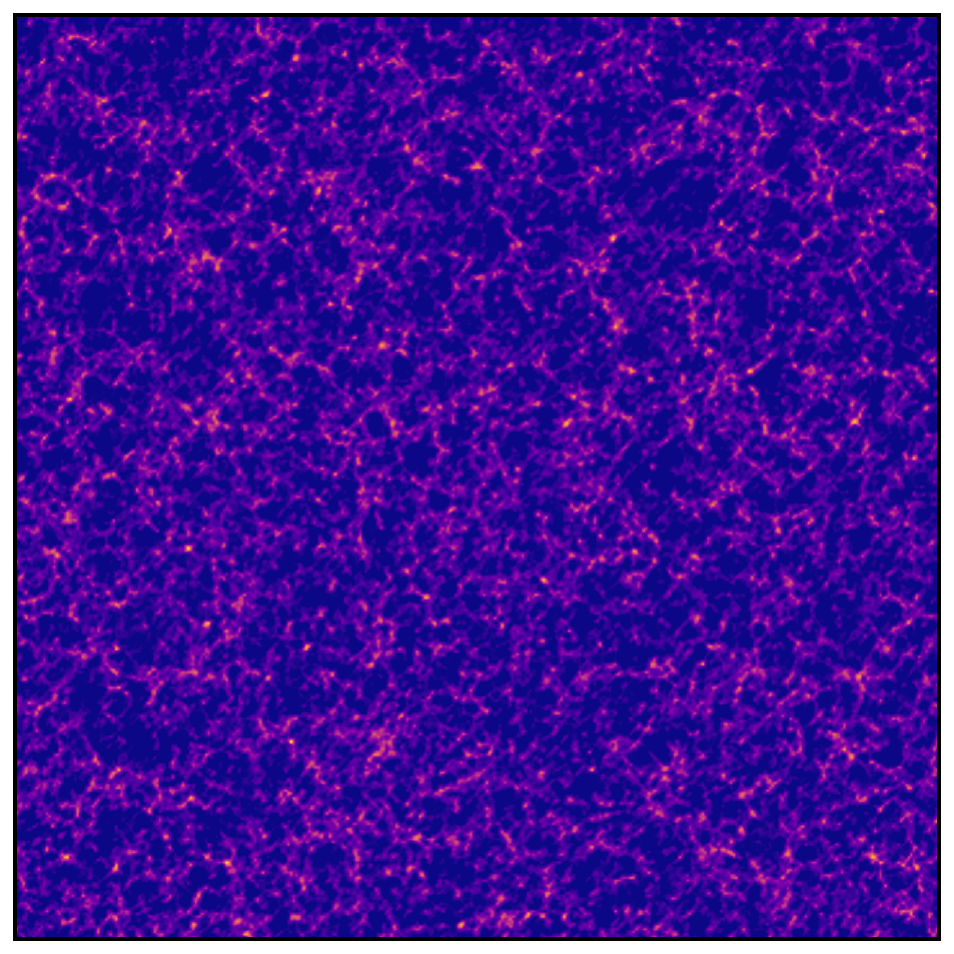}}\put(-218,207){\fcolorbox{white}{white}{\color{black}$-k^{0}$}}
\vspace{-0.4cm}
\\
\subfigure{\includegraphics[width=0.5\textwidth]{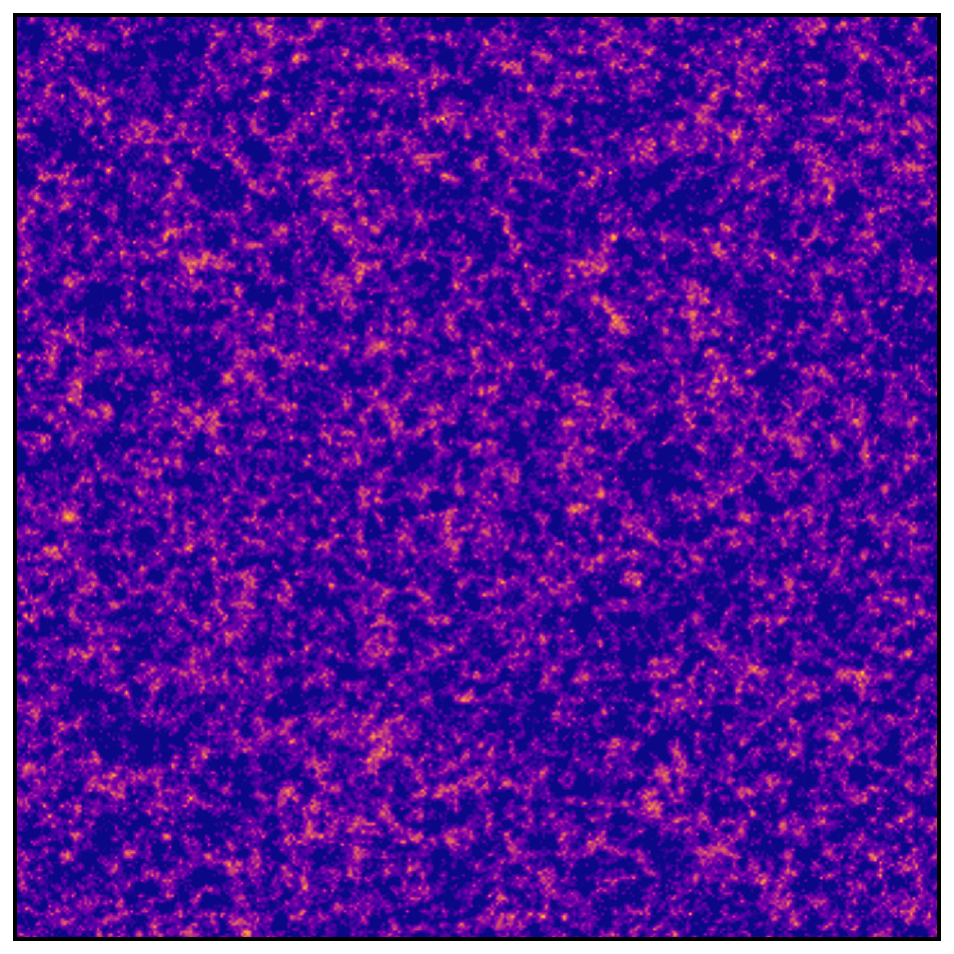}} \put(-218,207){\fcolorbox{white}{white}{\color{black}$-k^{2}$}} \hspace{-0.4cm} &
\subfigure{\includegraphics[width=0.5\textwidth]{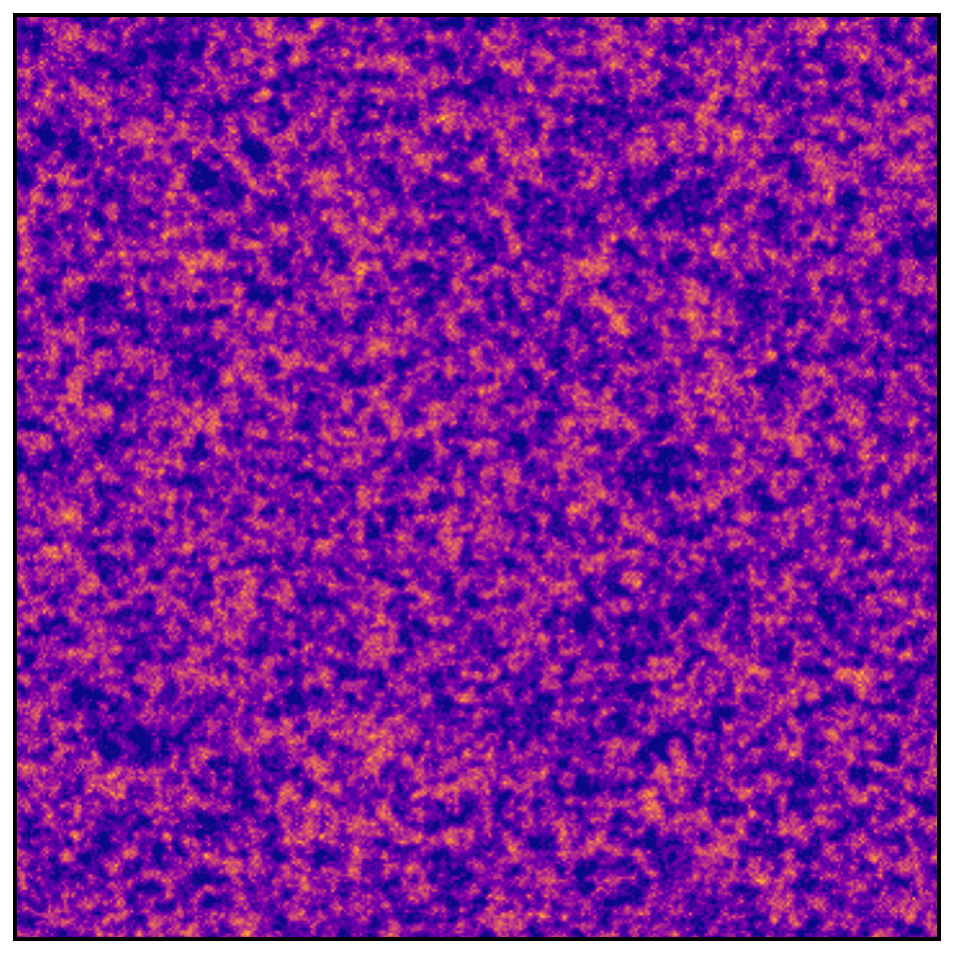}}\put(-218,207){\fcolorbox{white}{white}{\color{black}$-k^{4}$}}
\vspace{-0.4cm}
\end{tabular}
\put(-455,311){\rotatebox[]{90}{\large$\longrightarrow$}}    \put(-460,220){\rotatebox[]{90}{\large${\rm L}=2\,h^{-1}\,{\rm Gpc}$}}
\put(-455,115){\rotatebox[]{90}{\large$\longleftarrow$}}
\put(-400,328){\rotatebox[]{0}{\large$\longrightarrow$}}    \put(-380,328){\rotatebox[]{0}{\large${\rm d}L=5.55\,h^{-1}\,{\rm Mpc}$}}
\put(-283,328){\rotatebox[]{0}{\large$\longleftarrow$}}
\caption{{\bf Top left}: \textsc{Abacus} ``000'' distinct halos at $z=1.1$ assigned to a mesh. The side length is $2000\,h^{-1}{\rm Mpc}$ and the cell size is ${\rm d}L\simeq 5.55\,h^{-1}{\rm Mpc}$.
Other panels show the cosmic web classification contrast corresponding to the kernel hierarchy: ({\bf top right}) $-k^{-4}$,
({\bf middle left}) $-k^{-2}$, ({\bf middle right}) $-k^0$, ({\bf bottom left}) $-k^{2}$, ({\bf bottom right}) $-k^{4}$.}
\label{fig:cosmic-web}
\end{figure}
%=========================================================

%=========================================================
\begin{figure}[ht!]
    \centering
    \begin{tabular}{c}
    \includegraphics[scale=0.6]{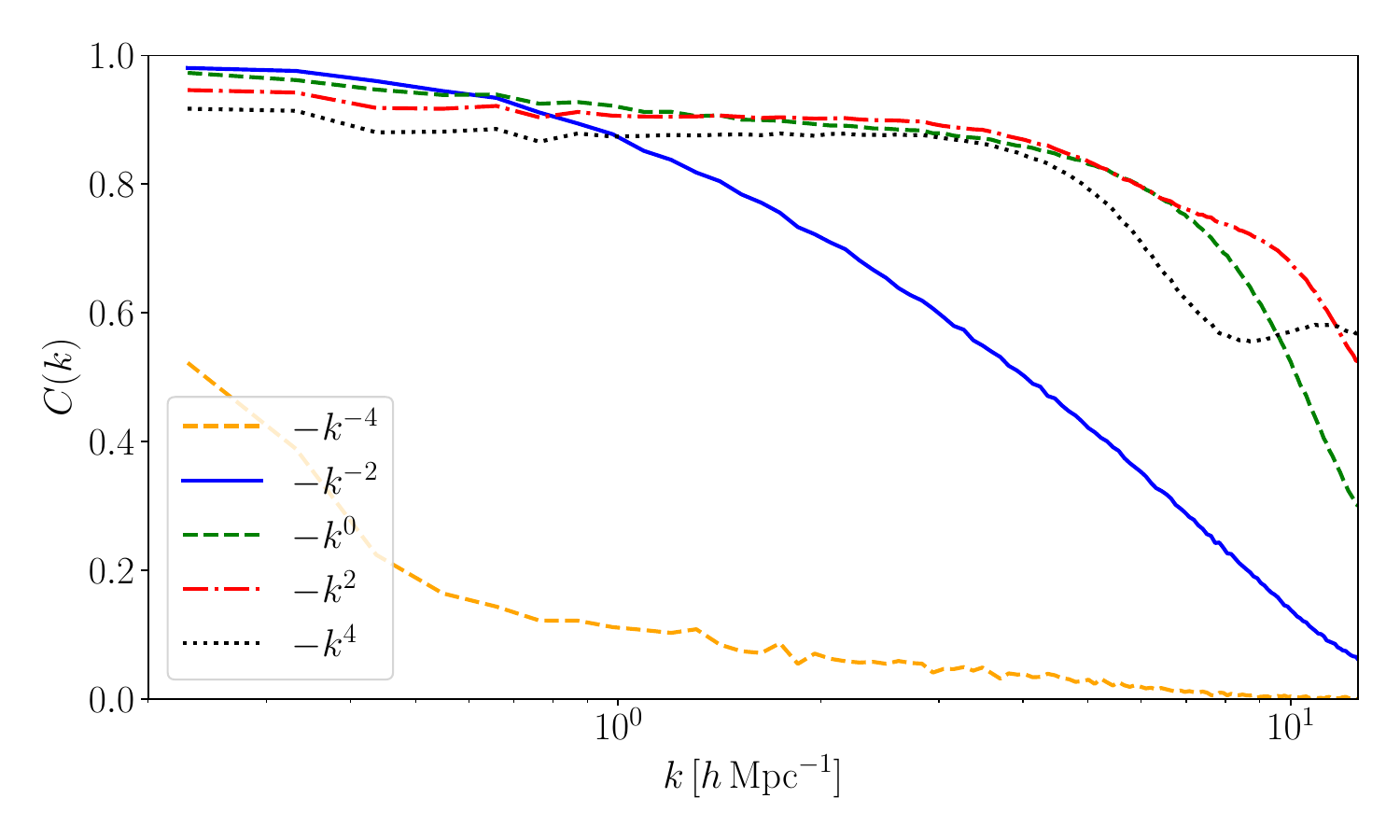}
    \put(-110,225){${\rm d}L=0.25\,h^{-1}\,\mathrm{Mpc}$}
    \put(-70,205){$z=1$}
    \put(-200,105){$\delta_\mathrm{web}^\mathrm{FastPM}\times\delta_\mathrm{DM}^\mathrm{FastPM}$}
    \vspace{-1.2cm}
           \\
    \includegraphics[scale=0.6]{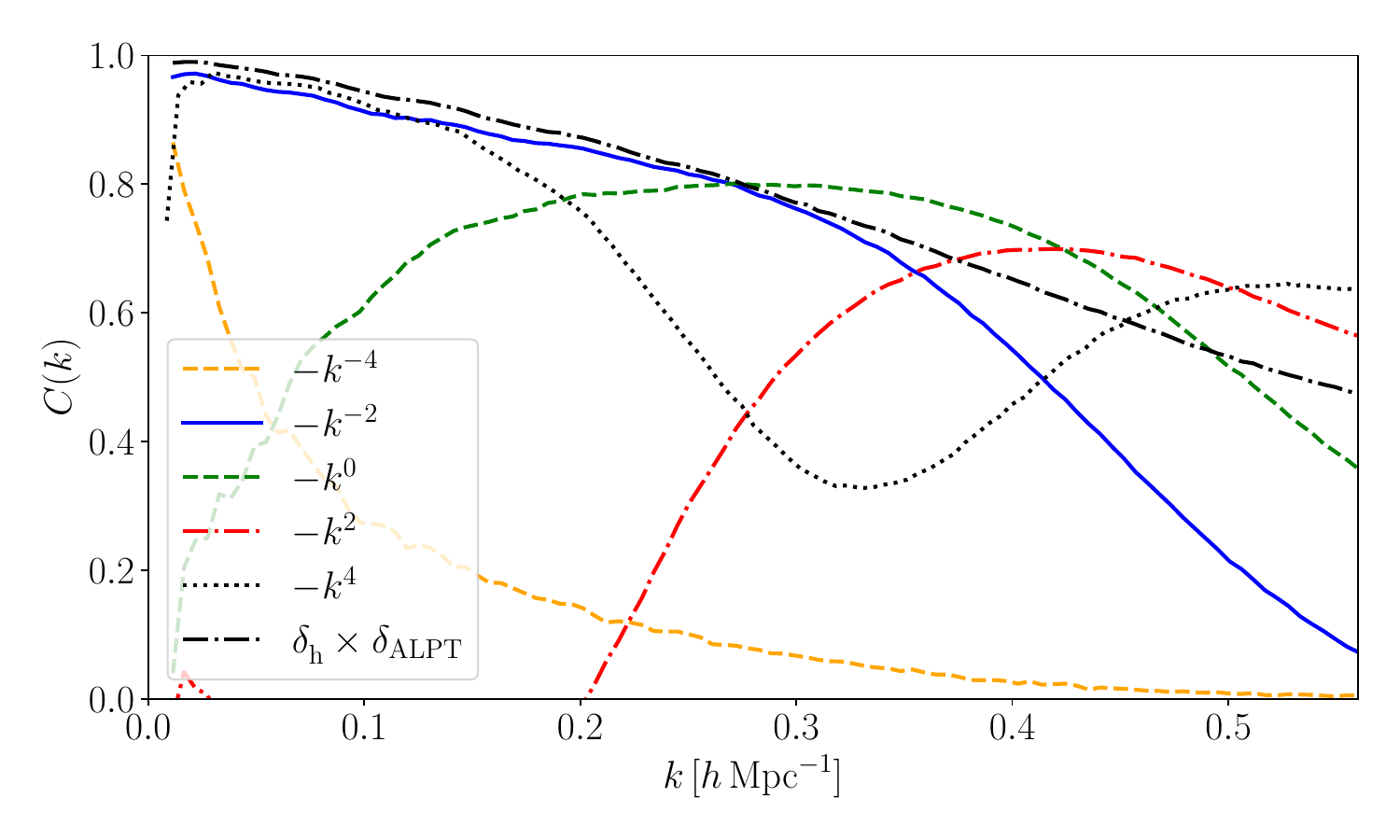}
    \put(-110,225){${\rm d}L=5.55\,h^{-1}\,\mathrm{Mpc}$}
    \put(-70,205){$z=1.1$}
    \put(-200,95){$\delta_\mathrm{web}^\mathrm{ALPT}\times\delta_\mathrm{h}^\mathrm{ABACUS}$}    \end{tabular}
    \vspace{-0.7cm}\caption{Cross-correlation coefficient $C^{(i)}(k)$ for the different hierarchy levels
    $i=-4,-2,0,2,4$.
    The coefficient is defined as
    $C^{(i)}(k)=\avg{\hat{\delta}^*_{\rm tr}(\bk)\hat{\delta}_{\rm web}^{(i)}(\bk)}/\sqrt{P_{\rm tr}(k)P_{\rm web}^{(i)}(k)}$,
    where $\delta_{\rm tr}$ denotes the tracer field and $\delta_{\rm web}^{(i)}$ the compressed web-density-contrast field at hierarchy level $i$.
    {\bf Top panel:} high-resolution \textsc{FastPM} matter field at $z=1$ with mesh spacing
    ${\rm d}L=0.25\,h^{-1}\,\mathrm{Mpc}$.
    {\bf Bottom panel:} \textsc{AbacusSummit} halo field at $z=1.1$, analysed on a coarse mesh with
    ${\rm d}L=5.55\,h^{-1}\,\mathrm{Mpc}$; the corresponding web fields are computed from an ALPT matter field generated from the down-sampled initial conditions of the same simulation.
    { The threshold is set to $\lambda_{\rm th}^{(i)}=0$ for all hierarchy levels except $i=4$, where a small stabilizing threshold $\lambda_{\rm th}^{(4)}=0.001$ is adopted. This numerical value is quoted in the eigenvalue units implied by our Fourier-space convention; since $\Hess^{(4)}_{ab}(\bk)\propto k_a k_b k^4\delta(\bk)$, the corresponding threshold carries the units of $k^6$ when physical wavenumbers are used and is therefore not directly comparable to thresholds at lower hierarchy levels. Its only purpose here is to regularise the most UV-sensitive classifier against mesh-scale fluctuations.}
    The figure shows that the classical tidal-web level $i=-2$ dominates on large scales, whereas the higher hierarchy levels
    $i=0,2,4$ retain significantly more information toward nonlinear scales, both in the high-resolution matter field and in the coarse-mesh halo analysis.
    }
    \label{fig:ck_dm_web}
\end{figure}
%=========================================================

%=========================================================

%=========================================================

%=========================================================
%=========================================================
\section{Discussion and conclusions}
\label{sec:concl}

We have introduced a spectral hierarchy of cosmic web classifications based on the kernel ladder
$\delta^{(i)}(\mathbf{k}) \equiv -k^i \hat{\delta}(\mathbf{k})$ for even integer $i$.
This construction unifies classical potential-based (tidal) and curvature-based web definitions,
connects naturally to long- and short-range nonlocal bias operators,
and extends the web concept to higher-derivative levels aligned with renormalised bias theory and EFT intuition.

The hierarchy provides a physically organized ladder of environmental descriptors.
At the long-range end ($i=-4,-2$), the operators probe infrared structure and tidal anisotropy.
At the intermediate level ($i=0$), curvature and peak morphology become dominant.
At higher levels ($i=2,4$), the hierarchy increasingly emphasizes short-range structure and higher-derivative operator content.
This ordering is not merely formal:
it manifests clearly both in configuration space (Fig.~\ref{fig:cosmic-web})
and in Fourier space (Fig.~\ref{fig:ck_dm_web}),
where the cross-correlation with halos shifts toward higher $k$ for increasing $i$.

A central conceptual point, inspired by the hierarchical cosmic web assembly-bias framework
\citep{Kitaura_2022,Coloma_2024},
is that cosmic web classification encodes the same tensor invariants that appear in nonlocal bias expansions.
The $\phi$-web ($i=-2$) implicitly conditions on $(\delta,s^2,s^3)$ through the invariants $(I_1,I_2,I_3)$,
while the curvature web ($i=0$) and higher-derivative levels ($i=2,4$)
encode short-range operator content related to density-derivative tensors.
The spectral hierarchy therefore provides a direct bridge between environmental segmentation
and the operator language of perturbative bias.

An important practical advantage of the hierarchical web approach
is that it avoids the pitfalls of explicitly truncated nonlocal bias expansions.
Rather than introducing many operator terms with free coefficients,
the tracer catalogue can be subdivided into web-conditioned subsets.
Each subset behaves like a separate tracer population,
analogous to mass binning but now organized by physically meaningful nonlocal information.
Local positive-definite bias models can then be applied within each subset.
This strategy naturally preserves positivity, limits parameter proliferation,
and maintains interpretability \citep[see][]{Coloma_2024,Sinigaglia_2024}.

Our information-theoretic test demonstrates that even a highly compressed
four-value web field retains significant correlation with the halo distribution
from $k\sim10^{-2}\,h\,\mathrm{Mpc}^{-1}$ up to the Nyquist scale of a
$\Delta x\simeq5.5\,h^{-1}\mathrm{Mpc}$ mesh.
The persistence of correlation at high $k$ is particularly noteworthy,
since the web field discards continuous density information
and preserves only categorical morphology.
The resulting 256-region PDF footprint (from four hierarchy levels)
provides a compact tracer signature that may serve as a practical basis
for bias learning and environmental characterization.

These findings have direct implications for fast mock galaxy catalogues.
The relevant $k$-range coincides with the resolution regime typical of
large-volume forward models and approximate gravity solvers.
The spectral hierarchy therefore offers a computationally cheap yet physically grounded
conditioning basis for mock generation.
It can be used to learn bias relations from high-fidelity reference catalogues \citep[e.g.,][]{2025arXiv251204362F}
and transfer them to approximate solvers operating on coarse meshes,
as in emerging bias-learning pipelines.
Because the hierarchy is discrete and low-dimensional,
it is inexpensive to compute and store on lightcones,
and can be incorporated either in parametric region-wise fits
or as structured features in controlled learning schemes.

Although the spectral hierarchy introduced here is not a fractal model,
it is naturally related to fractal descriptions of large-scale structure
in the sense that both characterize how morphology and clustering vary with scale.
The present framework organizes this scale dependence through Fourier-space kernel weighting,
whereas fractal approaches quantify it through real-space scaling exponents or dimensions.
In approximately scale-free regimes, both descriptions are connected through the scaling relations
between correlation functions and power spectra.

Beyond mock generation, the hierarchy provides a systematic framework
for studying galaxy evolution and morphology--environment relations.
By separating contributions from long-range tidal structure ($i=-2$),
curvature and peak morphology ($i=0$),
and higher-derivative local environment ($i=2,4$),
it becomes possible to disentangle environmental effects across physical scales.
This scale-organized environmental basis may prove useful
for investigating assembly bias, spin alignment, and morphology-dependent clustering
\citep[see, e.g.,][]{2014MNRAS.445..988N,2018MNRAS.476.3631P}.
This can be especially relevant in field-level inference \citep[see][]{2025arXiv250603969R}.

{
The spectral hierarchy can be extended in several directions.
Possible developments include:
\begin{itemize}
\item exploring the spectral hierarchy as a halo-identification framework beyond simple density maxima, complementing traditional density-based halo finders such as \textsc{Amiga Halo Finder} \citep[][]{Knollmann_Knebe_2009} and similar algorithms, by identifying relaxed structures through their multi-scale curvature and higher-derivative signatures;

\item incorporating velocity-shear analogues and connections to velocity bias, since velocity-field web classifications encode information related to the velocity divergence $\theta$ (with { $\theta=-\mathcal H f\delta$ at linear order, with $f=d\ln D/d\ln a$}), and have been shown to help capture halo bias information toward smaller scales \citep[see e.g.][]{Kitaura_2022};

\item optimizing eigenvalue thresholds $\lambda_{\rm th}^{(i)}$ using information criteria, thereby significantly reducing the arbitrariness in the choice of the threshold once one moves away from the commonly adopted $\lambda_{\rm th}=0$;

\item integrating hierarchy levels as conditioning variables in large-scale hierarchical assembly-bias models;

\item mapping alternative gravity models onto $\Lambda$CDM simulations \citep[see][]{2024A&A...690A..27G};

\item exploring multi-tracer applications in which different galaxy populations respond to different hierarchy levels, shedding light on the connection between cosmological large-scale structure and galaxy formation.
\end{itemize}
}

{ We stress that the revisions to the theoretical interpretation introduced above do not modify the operational definition or numerical implementation of the spectral cosmic-web classification, and therefore do not affect any of the classifications, cross-correlations, PDFs, or numerical results presented in this work. This distinction is important because the same spectral hierarchical classification is already being used in independent applications beyond the present manuscript: in a manuscript submitted to \emph{Physical Review Letters} on primordial non-Gaussianity, where environment-dependent responses of the hierarchy are exploited to model the large-scale PNG signal, and in a separate work in preparation applying the same framework to brain imaging for brain-age determination. These applications rely on the classification itself, which remains unchanged; the present revisions clarify only the physical interpretation and derivative bookkeeping of selected hierarchy levels.}

In summary, the spectral hierarchy provides both a conceptual unification
of cosmic web classifiers and a practical environmental compression scheme.
It organizes nonlocal information in a scale-ordered ladder,
demonstrably retains halo-relevant information up to mesh Nyquist scales,
and offers a structured, interpretable basis for next-generation mock generation
and environmental studies of galaxy formation.

%=========================================================

\section*{Acknowledgements}

{ The authors thank Pau Amaro-Seoane for a careful critical reading of the manuscript and for pointing out several important theoretical and presentational issues, in particular concerning the transformation of the effective growth equation, the derivative counting of the spectral hierarchy, and the interpretation of the ultraviolet hierarchy levels.}

The authors thank the \texttt{HOLI-mocks} team for useful discussions and testing the \textsc{WebON}-code, special mention to Ana Almeida, José María Coloma-Nadal, Ginevra Favole, Daniel Forero Sánchez, Jorge García Farieta, Fernando Frías García Lago, Pere Rosselló Truyols,  Yunyi Tang, Natalia Villa Nova Rodrigues, Cheng Zhao.
FSK acknowledges the Instituto de Astrof\'isica de Canarias (IAC) for continuous support to the \textit{Cosmology with LSS probes} activities
and the Spanish Ministry for Innovation and Science / Agencia Estatal de Investigaci\'on for support of the project
\textit{Big Data of the Cosmic Web} (PID2020-120612GB-I00/AEI/10.13039/501100011033), under which this work has been conceived and carried out, and the \textit{FIRE (Field-level bayesian Inference to Reconstruct the univErse)} project, which has been recently funded by a \textit{Proyecto de Generaci\'on de Conocimiento 2024}, PID2024-160504NB-I00.
FS acknowledges support from the \textit{Institute for Fundamental Physics of the Universe} postdoctoral fellowship scheme. FS is also grateful to IAC for hospitality during part of the realization of this project.

\section*{Disclaimer}
The \texttt{WebOn} code used in this work has been mainly written by FSK in C++ without AI.

\texttt{WebOn} is a modular framework whose methodological components are presented across three companion papers: the present work on the spectral hierarchical cosmic-web; a dedicated paper on ridged Lagrangian perturbation theory \citep{KitauraSinigaglia_RLPT_2026}; a further paper describing the lightcone evolution mode within the \texttt{HOLI-mocks} pipeline {to produce the DESI covariance mocks used for the DR2 Key Projects} \citep{KitauraSinigaglia_DESI_2026} (see: \url{www.cosmic-signal.org}).
\texttt{WebOn} has already been used in three previous publications: one presenting the \emph{Hierarchical cosmic web and assembly bias} \citep{Coloma_2024} approach for effective field-level bias modelling exploiting the hierarchical cosmic web $k^{-2}$ and $k^0$ levels; one presenting \emph{Cosmomia: cosmic-web based redshift-space halo distribution} \citep{Forero_2024} for subgrid modelling exploiting the knots $k^0$  level as attractors; and \emph{Fast and accurate Gaia-unWISE quasar mock catalogs from LPT and Eulerian bias} \citep{2025arXiv250915890S}, using the  two-hundred 10 $h^{-1}\,\mathrm{Gpc}$ side  cubical volume simulations  performed with the lightcone evolution \textsc{WebOn} code on the Leonardo@CINECA pre-exascale supercomputing facilities in September 2023 through the INAF\_C9A09 HPC proposal (637k core hrs). It has also served as a reference for its \texttt{JaX} differentiable version in \emph{Differentiable Fuzzy Cosmic-Web for Field Level Inference} \citep[][]{2025arXiv250603969R}, which is {under active development}.

%=========================================================
\bibliographystyle{JHEP}
\bibliography{references}

\end{document}